\documentclass[letterpaper,journal]{IEEEtran}
 
\usepackage{amsmath,amsfonts}
\usepackage{algorithmic}
\usepackage{algorithm}
\usepackage{array}
\usepackage[caption=false,font=normalsize,labelfont=sf,textfont=sf]{subfig}
\usepackage{textcomp}
\usepackage{stfloats}
\usepackage{cuted}
\usepackage{url}
\usepackage{verbatim}
\usepackage{graphicx}
\usepackage{cite}
\usepackage[export]{adjustbox}
\usepackage{xfrac}  

\usepackage{booktabs}
\usepackage{enumitem}

\usepackage{xcolor}
\usepackage[switch]{lineno}
\usepackage{lipsum}

\usepackage{amssymb}
\usepackage{booktabs}
\usepackage[bb=boondox]{mathalfa}

\usepackage[english]{babel} 
\usepackage{cancel}

\ifdefined\draft

    \usepackage{marginnote}
    \newcommand{\about}[1]{\marginnote{\textsf{\color{green}\scriptsize#1}}}
    
\else
    \newcommand{\about}[1]{}
\fi

\ifdefined\draft
    \newcommand{\todo}[1]{\textsf{\color{red}\scriptsize[#1]}}
\else
    \newcommand{\todo}[1]{}

\fi

\let\oldnl\nl
\newcommand{\nonl}{\renewcommand{\nl}{\let\nl\oldnl}}

\usepackage{wrapfig}
\usepackage{blindtext}
\usepackage{upgreek}

\newcommand{\eq}[1]{\eqref{eq:#1}}
\newcommand{\fig}[1]{Figure~\ref{fig:#1}}
\newcommand{\tab}[1]{Table~\ref{tab:#1}}

\newcommand{\sctn}[1]{Section~\ref{sec:#1}}
\newcommand{\appx}[1]{Appendix~\ref{app:#1}}

\newcommand{\n}[1]{\mathrm{#1}} 
\newcommand{\mb}[1]{\mathbf{#1}}
\newcommand{\mbs}[1]{\boldsymbol{#1}}
\newcommand{\mbb}[1]{\mathbb{#1}}
\newcommand{\mc}[1]{\mathcal{#1}}

\newcommand{\diag}{\mathop{\mathrm{diag}}}

\DeclareMathOperator*{\argmin}{arg\,min}

\begin{document}

\ifdefined\draft
    
\setcounter{page}{0}
\textbf{EDICS}

\begin{itemize}
    \item \textbf{CIF-OBI}	
    \item \textbf{CIS-SSI}	
    \item \textbf{CIS-TIM}	
\end{itemize}

\vspace{1cm}

{\adjincludegraphics[width=0.8\columnwidth,
trim={{0.0\width} {0.0\height} {0.0\width} {0.0\height}}
,clip]{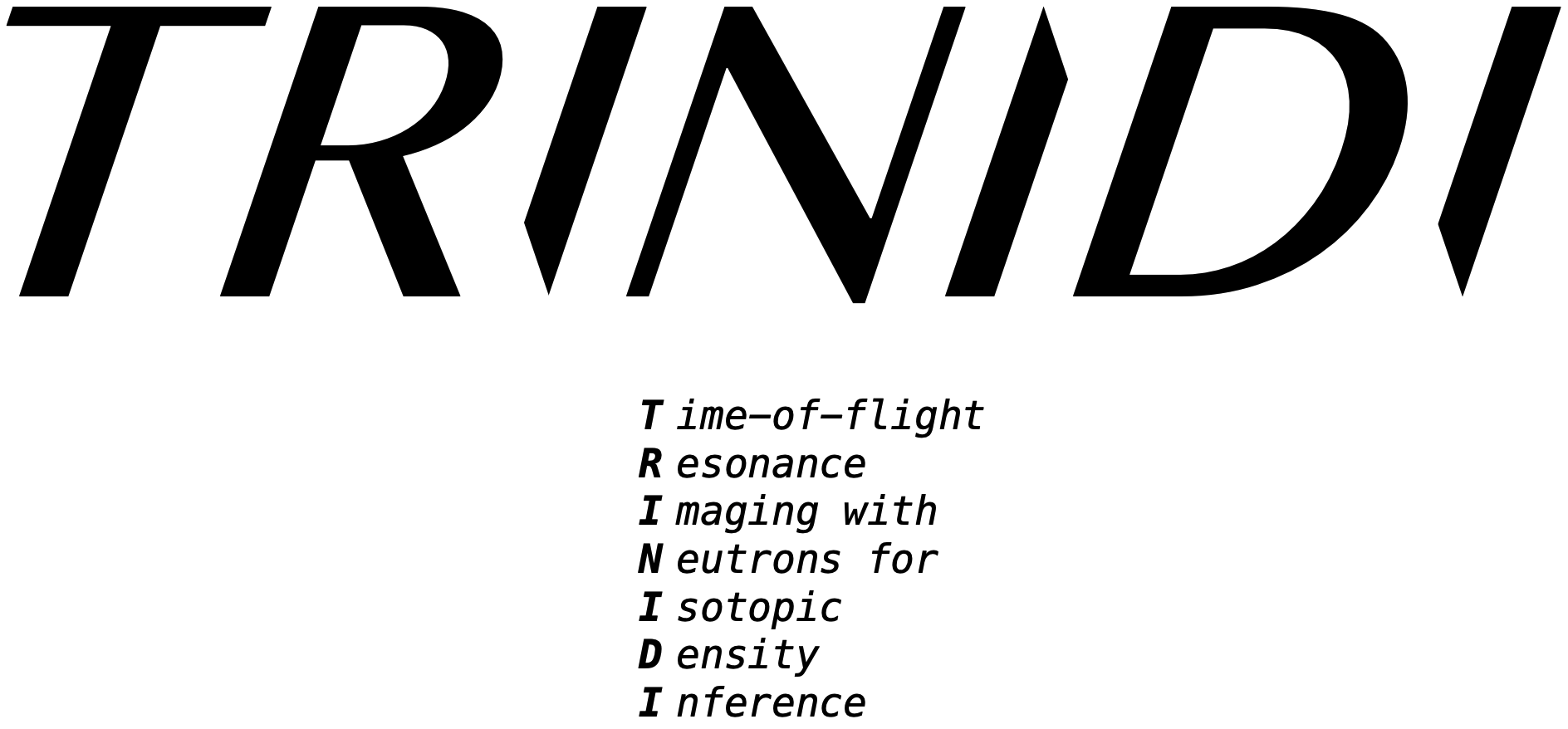}}

\vspace{1cm}

\texttt{LA-UR-23-21022}

\todo{ToDo}


\fi

\title{TRINIDI: Time-of-Flight Resonance Imaging with Neutrons for Isotopic Density Inference}

\author{
    Thilo Balke,~\IEEEmembership{Student Member,~IEEE},
    Alexander M. Long,
    Sven C. Vogel,
    Brendt Wohlberg,~\IEEEmembership{Fellow,~IEEE},
    Charles A. Bouman,~\IEEEmembership{Fellow,~IEEE} 
    \thanks{
        Thilo Balke and Brendt Wohlberg are with Theoretical Division, Los Alamos National Laboratory, Los Alamos, NM 87545 USA. 
        E-mail: \texttt{thilo.balke@gmail.com}, \texttt{brendt@lanl.gov}
    }
    \thanks{
        Alexander M. Long and Sven C. Vogel are with Materials Science and Technology Division, Los Alamos National Laboratory, Los Alamos, NM 87545 USA. E-mail: \texttt{alexlong@lanl.gov}, \texttt{sven@lanl.gov}
    }
    \thanks{
        Charles A. Bouman is with the School of ECE and BME of Purdue University, West Lafayette, IN 47907 USA. E-mail: \texttt{bouman@purdue.edu}
    }
    \thanks{
        Research was supported by the Laboratory Directed Research and Development program of Los Alamos National Laboratory under project number 20200061DR, and partially supported by NSF grant number CCF-1763896.
    }
}



\maketitle

\begin{abstract}

\about{problem formulation}
Accurate reconstruction of 2D and 3D isotope densities is a desired capability with great potential impact in applications such as evaluation and development of next-generation nuclear fuels.
Neutron time-of-flight (TOF) resonance imaging offers a potential approach by exploiting the characteristic neutron absorption spectra of each isotope.
However, it is a major challenge to compute quantitatively accurate images due to a variety of confounding effects such as severe Poisson noise, background scatter, beam non-uniformity, absorption non-linearity, and extended source pulse duration.

\about{our contribution}
We present the TRINIDI algorithm which is based on a two-step process in which we first estimate the neutron flux and background counts, and then reconstruct the areal densities of each isotope and pixel.
Both components are based on the inversion of a forward model that accounts for the highly non-linear absorption, energy-dependent emission profile, and Poisson noise, while also modeling the substantial spatio-temporal variation of the background and flux. 
To do this, we formulate the non-linear inverse problem as two optimization problems that are solved in sequence.
We demonstrate on both synthetic and measured data that TRINIDI can reconstruct quantitatively accurate 2D views of isotopic areal density that can then be reconstructed into quantitatively accurate 3D volumes of isotopic volumetric density.

\end{abstract}


\section{Introduction}
\label{sec:intro}

\about{general neutron imaging}
\IEEEPARstart{N}{eutrons}, unlike X-rays, protons, or electrons, primarily interact with the nucleus rather than the electron shell of atoms, and thus neutron radiography can offer complementary contrast mechanisms to more conventional radiography probes~\cite{banhart2010x}.
While X-rays generally get attenuated more by heavier isotopes, this simple relationship does not hold for neutron attenuation.
Neutron imaging thus enables the non-destructive evaluation of nuclear fuel, bulk objects, and other materials formed of heavy nuclei~\cite{anderson2009neutron}, while revealing (X-ray-transparent) features such as plant roots and soil water~\cite{tumlinson2008thermal, malone2016vivo}. 
Applications range from direct methods such as transmission radiography and tomography as well as indirect methods that leverage neutron diffraction and small angle scattering to determine crystal structures, shapes of biological molecules as well as atomic resolution techniques such as neutron holography~\cite{vontobel2006neutron, anderson2009neutron, kardjilov2018advances}.
In addition, simultaneous X-ray and neutron tomography can provide multi-modal attenuation mapping such as the mechanical degradation (via X-rays) as well as chemical diffusion process (via neutrons) in lithium batteries~\cite{ziesche20204d, tengattini2020next}.

\about{neutron TOF imaging}
Neutron time-of-flight (TOF) imaging is a promising imaging technique that exploits the property that neutrons travel with a velocity that depends on their energy.
At advanced neutron source facilities, such as ISIS~\cite{thomason2019isis}, LANSCE~\cite{lisowski2006alamos}, SNS~\cite{bilheux2014neutron}, and J-PARC~\cite{ikeda2005current} spallation neutrons are pulsed and moderated down to yield a broad spectrum of energies. These types of neutron sources allow for neutron TOF measurements, where neutron energies are determined by the time it takes to traverse a known distance between the source and the detector.
Thus, pulsed neutron TOF measurements are inherently hyperspectral, and can be used to measure neutron absorption or diffraction as a function of energy~\cite{LEHMANN2009429, Hilger:15, kockelmann2007energy, schillebeeckx2014neutron} facilitating, for example, 2D temperature mapping \cite{sato2009pulsed, tremsin2015spatially} through Doppler broadening of neutron absorption resonances as well as neutron diffraction imaging techniques for identification of crystal structure and material strains~\cite{tran2021spectral, woracek2011neutron}. 

\about{problem: isotopic densities\\from transmission TOF data}
This paper deals with the problem of reconstructing isotopic densities from neutron TOF transmission measurements given the knowledge of isotope-specific neutron cross section spectra.
The neutron energies suitable for this modality range from approximately $1 \, \n{eV}$ to $10 \, \n{k eV}$, categorized as resonance neutrons~\cite{schillebeeckx2014neutron}.
Absorption resonances are isotope specific and increase the attenuation of neutrons by orders of magnitude within a narrow energy range~\cite{breit1936capture}.
Quantitatively accurate isotope reconstruction is very challenging for a number of reasons. 
TOF neutron imaging measurements are typically exceptionally noisy, with average neutron counts as low as only $2$ counts per measurement bin after several hours of measuring.
This is mainly because each measurement bin only corresponds to a single small pixel ($\lesssim 100 \, \upmu$m) and a short TOF window ($\lesssim 100 \, \n{ns}$) and because even the most advanced neutron sources emit several orders of magnitude fewer particles than typical X-rays sources~\cite{MOCKO2008455, ino2004measurement}. 
In addition, the measurements contain substantial background counts which are typically about as strong as the signal itself~\cite{tremsin2013non}. 
Finally, the measurement physics are highly non-linear due to the non-linear absorption, the large dynamic range of the spectral neutron responses of the nuclei, and the effects of the extended pulse duration of the neutron beam, an energy-dependent phenomenon that is caused by the moderation process\cite{schillebeeckx2014neutron}.

\about{low noise single spectra}
The usual method for low-noise measurements~\cite{postma2009neutron, schillebeeckx2012determination, ma2020non, hasemi2015evaluation} is to first compute a transmission spectrum (i.e. open beam normalized transmission measurement) from the TOF data and then to process it with software packages such as SAMMY~\cite{Larson2008}, CONRAD~\cite{de2007status} or REFIT~\cite{moxon1991least}. 
However, these software packages require relatively low-noise estimates of the transmission spectrum as input. 
In addition, since these software packages are primarily designed for the analysis of individual neutron spectra, they would be very slow when applied to a large array of pixels.

\about{imaging crude methods}
In an imaging application low-noise spectra can be obtained by selecting and averaging regions of interest which can then be processed as a 1D signal~\cite{tremsin2017nond, tremsin2013non}. However, the averaging does not yield density estimates on a per-pixel basis, so the problem is shifted to the accurate estimation of transmission spectra at each pixel.
It is due to this difficulty that the vast majority of papers on neutron resonance imaging is limited  to 2D mapping of isotopes rather than quantification of isotope densities. Our work here aims to overcome these shortcomings.

\about{difficulty computing transmission}
The transmission spectrum is the expected ratio of the sample and open beam scans, both of which must be corrected for the severe background and beam non-uniformity. 
There are a number of reasons for the difficulty of accurately estimating these transmission spectra.
Although explicit models of the background have been proposed~\cite{schillebeeckx2012determination}, there are no widely accepted methods for jointly estimating background and beam non-uniformity.
Moreover, these problems are exacerbated by the need to separately estimate the transmission at each pixel.
This means that the spatio-temporal non-uniformity in the background and beam-profile must be accounted for along with the very high level of neutron counting noise.
Importantly, since the transmission spectrum estimate is a ratio of two very noisy signals, the resulting spectrum is even noisier.

\about{2D, 3D densities}
Notably among the early efforts of estimating 2D and 3D isotope densities are the work of Sato, Kamiyama, et al. \cite{sato2009pulsed, kamiyama2009energy}. Here a large $8 \!\times\! 8$ pixel-coarse imaging detector is used to yield low-noise measurements. In addition, the effects of background events were neglected. In \cite{festa2015neutron} a similarly sized detector was used where the estimation was based on the assumption that the isotope densities are simply proportional to the area of the transmission resonance dips. Losko, Vogel et al.~\cite{losko20223d, vogel2018neutron} accomplished density estimates with a pixel-dense imaging detector. Similar to the previous cases, the transmission spectra are relatively low noise likely due to a combination of much larger TOF bins or spatial denoising pre-processing. Such measures limit spatial and temporal resolution while allowing for only restricted energy range to only about $10 \, \n{eV}$. Our work here focuses on low count, high noise measurements.

\about{our previous work}
Previously we proposed a method~\cite{balke2021hyperspectral} based on a linearized model and the assumption of opaque ``black resonances''~\cite{syme1982black} to estimate the background counts.
While this research demonstrates fast and semi-quantitative results, it has a number of limitations.
First, black resonances are not always available or entirely opaque after the blur caused by the extended pulse shape (see \appx{ResolutionFunction}).
Second, the linearized model may not be sufficiently accurate as it neglects to model the extended pulse duration and model the counting noise adequately.
Third, this approach implicitly still first estimates the very noisy transmission spectra and as a consequence densities of highly attenuating isotopes can be strongly under-estimated. (See the $^{238}\n{U}$ and $^{240}\n{Pu}$ estimates in Table 2 of \cite{balke2021hyperspectral}.)

\about{this paper}
In this paper, we present the Time-of-flight Resonance Imaging with Neutrons for Isotopic Density Inference (TRINIDI) algorithm. An open-source Python implementation~\cite{trinidi} is available in the spirit of reproducible research.
Our model-based iterative reconstruction (MBIR~\cite{buzzard2018plug, balke2018separable, majee2021multi, Venkatakrishnan2013pnp}) algorithm is based on a two-step process in which we first estimate the neutron flux and background, and then estimate the areal density of each isotope. Importantly, the initial computation of transmission spectra is avoided by modeling the neutron measurement in their native counting domain.
Both components are based on the inversion of a forward model that accounts for the highly non-linear absorption, extended pulse duration, and Poisson noise, while also modeling the very substantial space and energy varying background. 
To do this, we formulate the non-linear inverse problem as two optimization problems that are solved in sequence.

\about{results}
We perform numerical experiments on both simulated and measured data that demonstrate the ability of TRINIDI to reconstruct quantitatively accurate isotopic areal density from 2D data. 
Finally, we use a series of 2D views of a nuclear fuel sample that were experimentally measured at the LANSCE facility~\cite{lisowski2006alamos} to compute a quantitatively accurate 3D reconstruction of the fuel sample's material composition.

\section{The Neutron Imaging System }
\label{sec:ImagingSystem}

\begin{figure}[htb]
\vspace{-0.09cm} 
\centering
{\adjincludegraphics[width=1.0\columnwidth,
trim={{0.31\width} {0.46\height} {0.29\width} {0.32\height}}
,clip]{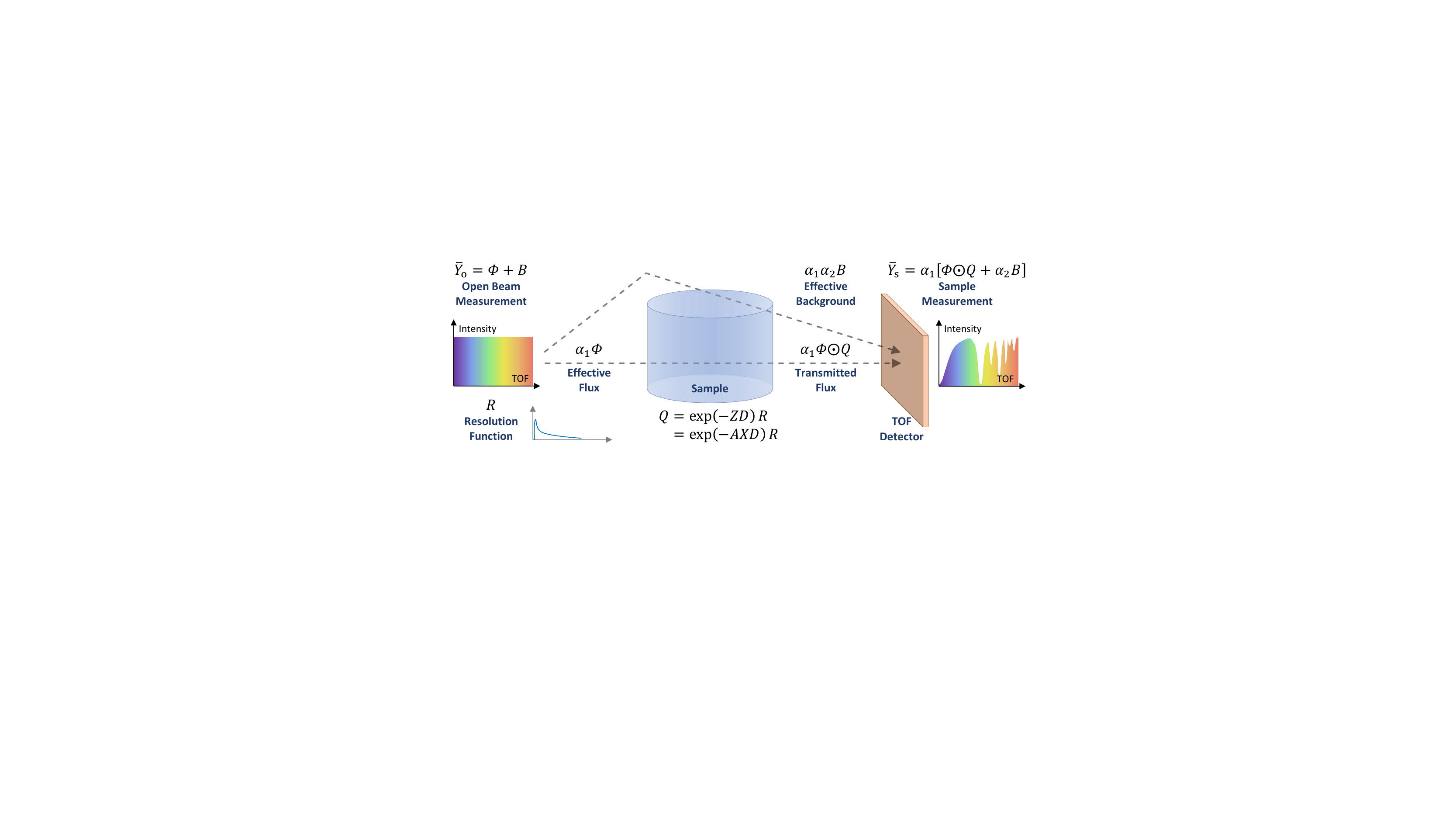}}
\caption{Illustration of the neutron flight path. A neutron pulse is emitted from the source, attenuated as it passes through the sample, and then detected by a high-speed camera. The time-of-flight (TOF) is used to separate neutrons into a large number of spectral components. The background is illustrated as a jagged line since scattering is assumed to be one source of background.}
\label{fig:flightpath}
\vspace{-0.09cm} 
\end{figure} 

\about{fightpath basics}
\fig{flightpath} illustrates the major components of a neutron beamline. Each pulse contains neutrons with a wide range of energies. The flux neutrons are (quasi-)exponentially attenuated as they pass through a sample, and finally, they are detected by a high speed imaging detector that collects approximately $2000$ equi-spaced frames at a rate of $30 \, \n{ns}$ per frame. The background events are also counted by the detector and are assumed to derive from a variety of scattering events in the flight path.

\about{TOF-energy correspondence}
Neutrons, unlike photons, have non-zero mass, so their energy is a function of their velocity, which is in turn a function of their TOF. Consequently, each TOF corresponds to a distinct neutron energy, and the set of images represent a spectral decomposition of the detected neutrons. The neutron velocities are sufficiently low for relativistic effects to be ignored, so the kinetic energy of the neutron is given by
\begin{equation}
\label{eq:energy_tof}
E = \frac{1}{2} m v^2 = \frac{1}{2} m \left( \frac{L}{t} \right)^2 \ ,
\end{equation}
where
$ m \approx 1.045 \!\times\! 10^{-8} \,  \n{eV \, s^2 / m^2}$
is the mass\footnote{We use $E=mc^2$ to express the unit of mass in terms of $\n{eV}$.}, $v$ the velocity, and $t$ the TOF of a neutron and $L$ is the flight path length.\footnote{This means that $t \propto \sfrac{1}{\sqrt{E}}$, and we use TOF and energy interchangeably}

\begin{figure}[htb]
\vspace{-0.09cm} 
\centering
{\adjincludegraphics[width=0.8\columnwidth,trim={{0.04\width} {0.2\height} {0.15\width} {0.15\height}},clip]{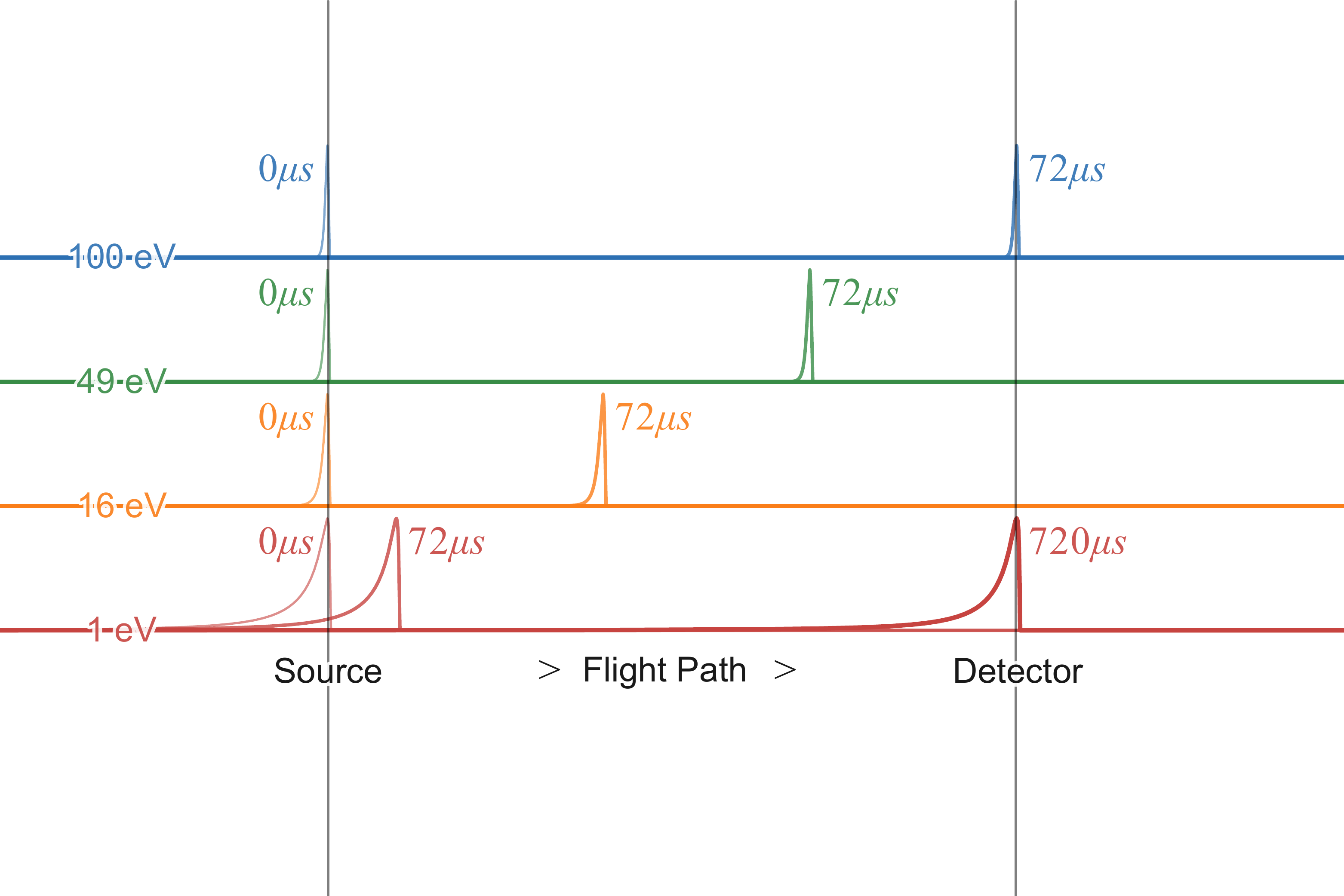}} 
\caption{Propagation of a neutron pulse through an $L = 10 \, \n{m}$ flight path. Higher energies correspond to shorter TOFs.
The pulse is energy-dependent.}
\label{fig:pulses}
\vspace{-0.09cm} 
\end{figure}

\fig{pulses} illustrates the propagation of a neutron pulse through an $L = 10 \, \n{m}$ flight path. For this distance, a $1 \, \n{eV}$ neutron corresponds to a TOF of $720 \, \n{\upmu s}$, and a $100 \, \n{eV}$ neutron corresponds to a TOF of $72 \, \n{\upmu s}$. Notice that the higher energy neutrons travel faster, and a $100\times $ increase in energy results in a $10\times$ decrease in TOF.

\about{pulse duration}
Notice that the initial pulse is not instantaneous. This means that neutrons with a particular energy may arrive over a range of TOFs. Equivalently, the arrival time does not precisely determine the energy of the neutron. A wider pulse has the effect of blurring absorption features in the time domain and thus limiting the temporal resolution of the imaging system\footnote{similar to a pinhole camera where a wider aperture causes spatial blurring.}. A longer flight path reduces the blurring\footnote{The absolute temporal blur is independent of the flight path length. However the longer the flight path the more the time-of-arrival is dominated by the velocity (energy) of the neutron as opposed to the uncertainty in the time the neutron enters the flight path. This effect reduces the relative blur observed at the detector plane.}, but also reduces the overall flux via the inverse-square law. There is therefore an inherent trade-off between temporal resolution and neutron flux.
Lastly, notice that the duration of the pulse is energy-dependent\footnote{The energy-dependence of the pulse length comes from the moderation process, a mechanism to create a wide spectrum from a high-energy pulse. Lower energy neutrons are emitted with a longer pulse since there is a larger number of possible collisions in the moderator to arrive at that lower energy.}. 
Accurate modeling of this effect will prove to be very important in solving our problem.

\about{energy range}
Our experiments will focus on neutron energies in the range of approximately $1 \, \n{eV}$ to $100 \, \n{eV}$, i.e. resonance region neutrons.
At resonances, the interaction probability of an incident neutron can increase by orders of magnitude and cause strong, isotope specific attenuation peaks. 
Below $1 \, \n{eV}$ isotopes usually do not have neutron resonances, and thus do not provide sufficient isotope-specific characteristics. 
At energies above about $100 \, \n{eV}$ the neutron resonances are so dense that the inherent temporal blur due to the neutron pulse renders individual resonances unresolvable. 
Furthermore, at such high energies, the neutron counts are dominated by background, which makes the very noisy signal less useful.

\begin{figure}[htb]
\vspace{-0.09cm} 
\centering
{\adjincludegraphics[width=1.0\columnwidth,
trim={{0.04\width} {0.04\height} {0.05\width} {0.11\height}}
,clip]{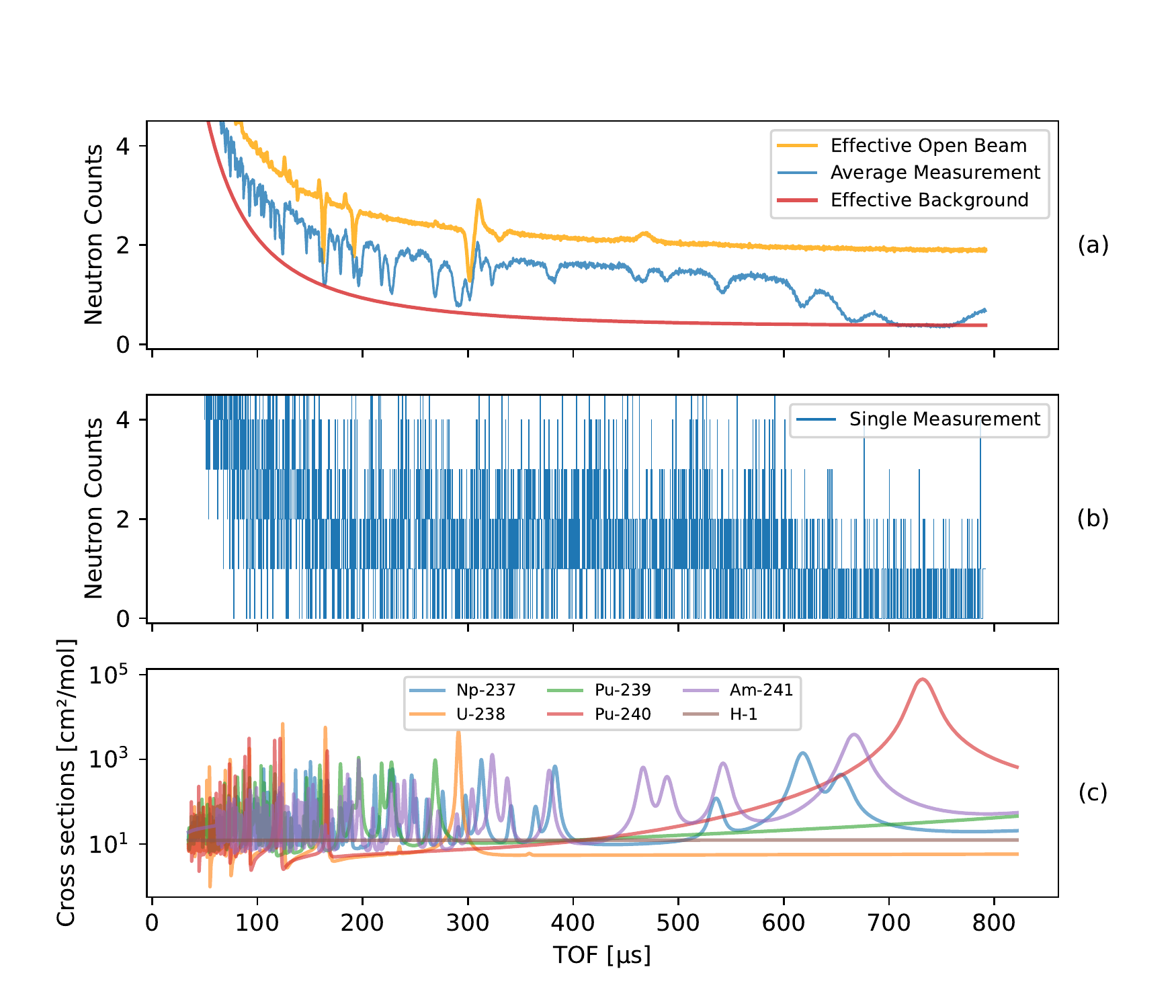}}
\caption{Spectral characteristics of neutron data as a function of TOF.\\
(a) Typical neutron measurement averaged over many pixels (blue), background neutron counts (red), open beam measurement (yellow)\\
(b) Neutron counts for a single pixel. \\
(c) Neutron cross sections for the isotopes in the sample. }
\label{fig:realspectra}
\vspace{-0.09cm} 
\end{figure}

\about{a typical measurement}
\fig{realspectra}(a) shows a neutron spectrum as a function of the TOF. 
The blue line shows the number of counts averaged over many pixels of the measured spectrum through a sample. 
Notice that the spectrum shows a great deal of fine structure, i.e. the neutron resonances, that can be used to distinguish different isotopes. 
Since neutrons interact with the nuclei, these characteristic resonances can be used to distinguish between different isotopes of the same element, such as the fissile $^{235}\n{U}$ and the non-fissile $^{238}\n{U}$ uranium isotopes. 
The red line shows the background counts which are quite large, and can result from a wide range of scattering effects in the flight path and sample. 
Finally, the yellow line shows the open beam measurement that results from a combination of the flux and the background events. 
Importantly, the background is large and the flux has a great deal of variation, so both must be accurately modeled in order to produce an accurate reconstruction of a sample.

\about{single pixel measurement}
\fig{realspectra}(b) shows a plot of a spectral neutron measurement of a single pixel. Since the measurement consists of neutron counts, it is discrete, and it is clear that for individual pixels the signal-to-noise (SNR) ratio is very low. Neutron TOF imaging is challenging due to the very low SNR.

\about{cross section spectra}
Finally, \fig{realspectra}(c) illustrates the cross section of various isotopes in the sample, in units of $\n{cm^2/mol}$, as a function of neutron TOF. The concept of a neutron cross section is used to express the likelihood of interaction between an incident neutron and a target nucleus. As the cross section of the material increases, it more strongly attenuates the neutrons and reduces the transmission. For the purposes of this research, these cross section spectra are considered known\footnote{Cross section spectra are tabulated in databases such as ENDF/B-VIII.0\cite{conlin2018release}, which are frequently updated through more accurate measurements.} for particular isotopes that make up the sample. 

\about{first goal:\\areal density}
As an example, imagine a sample made of a single isotope with a known cross section of $\mb{d}_j \, ^{[\n{cm^2/mol}]}$ at TOF bin $j$.
Furthermore, assume that the sample is of uniform thickness $\ell \, ^{[\n{cm}]}$,
with mass density $\rho \, ^{[\n{g/cm^3}]}$,
and molar mass $M \, ^{[\n{g/mol}]}$,
then it has a uniform areal density of
\begin{equation}
    \hspace{2.0cm}
z = \frac{\ell \rho}{M} \, .
\hspace{1.0cm} [\n{mol/cm^2}]
\label{eq:arealdensity_plate}
\end{equation}
Using Beer's Law, we would expect the transmission of the neutrons through the material at TOF bin, $j$, to be
$$
\mb{q}_j = \exp \left\{ - z \mb{d}_j \right\} \,.
$$
Our first goal will be to estimate the areal density, $z$, for each pixel and each material in the sample. 
The areal density is predominantly desired with a quasi-2D, flat sample such as a foil or a physical slice of a larger sample.

\about{second goal:\\ volumetric density}
Our second goal will be to estimate the volumetric density for each isotope in the sample. 
This is important when the sample is an arbitrarily shaped 3D object, such as the nuclear fuel sample used in our experiments.
In this case, our goal is to reconstruct the 3D volumetric density, $x$, where
\begin{equation}
    \hspace{2.0cm}
    x = \frac{\rho}{M} \, .
    \hspace{1.0cm}
    [\n{mol}/\n{cm}^3]
    \label{eq:volumetricdensity_plate}
\end{equation}
We can estimate $x$ for each voxel and material in the sample by extracting the areal densities for each rotational view followed by tomographic reconstruction.

\section{Single-Radiograph Forward Model}
\label{sec:ForwardModel}

\about{goal}
In this section, we derive our forward model for the hyperspectral neutron measurements of a radiograph. 
Our goal will be to estimate a matrix $Z \in \mbb{R}^{N_{\n{p}} \times N_{\n{m}}}$ where $N_{\n{p}}$ is the number of scalar projections through a sample (pixels) and $N_{\n{m}}$ is the number of unique isotopes that make up the sample.
Then $Z_{i,m}$ represents the areal density in units of $\n{mol}/\n{cm}^2$ of the $i^{\n{th}}$ projection through a sample for the $m^{\n{th}}$ isotope. 

\about{cross section dictionary}
The matrix $D \in \mbb{R}^{N_{\n{m}} \times N_{\n{F}}}$ denotes the cross section dictionary in units of $\n{cm}^2/\n{mol}$ for $N_{\n{F}}$ neutron energies. I.e. $D_{m,j}$ is the cross section for material $m$ at TOF $j$. Notice that each resonant isotope in the sample must be accounted for because distinct isotopes typically have distinct spectral signatures.

\about{initial model}
Assuming the conventional Beer-Lambert's law attenuation model, the expected measured neutron counts would be
\begin{equation}
\bar{Y}_{\n{s}} = \alpha_1 \left[ \varPhi \odot \exp \left\{- Z D \right\} + \alpha_2 B \right] \,,
\nonumber     
\end{equation}
where $\alpha_1$ and $\alpha_2$ are positive scalars, the symbol $\odot$ is the Hadamard product operator, $\varPhi_{i,j}$ is the neutron flux, and $B_{i,j}$ is the background for projection $i$ and TOF $j$, respectively.

\about{flux + background}
The flux, $\varPhi$, and background, $B$, are important scan parameters that vary as a function of position and energy.
In X-ray imaging, the background is relatively small and thus the flux can be accurately measured with an ``open beam'' calibration procedure.
However, in neutron imaging, the background tends to be quite large, somewhat sample dependent, and difficult to measure separately.
Consequently, it is also very difficult to measure the flux directly from a single open beam measurement.
Therefore, a major challenge will be to formulate an imaging methodology that allows us to recover both the flux and background from available measurements.

\about{resolution function}
From \sctn{ImagingSystem}, we know that the neutron TOF varies with energy.
However, in practice, the measured time of arrival (TOA) does not exactly determine the theoretical time of flight (TOF) of the neutron due to uncertainty in the time the neutron is emitted.
The neutrons are emitted in a causal pulse distribution with a long tail that has significant time duration as compared to the sampling rate of the detector camera. A neutron with a specific energy ($\leftrightarrow$ TOF) therefore has finite probability of being detected over a range of times ($\leftrightarrow$ TOA).

In order to model this effect, we introduce the resolution function or resolution operator, $R\in \mbb{R}^{N_{\n{F}} \times N_{\n{A}}}$, representing the conditional probability
\begin{equation}
\label{eq:resolution_def}
   R_{i,j} = \mbb{P}\left\{ \mbox{TOA} = \mb{t}^{\n{A}}_j 
   \ | \
   \mbox{TOF} = \mb{t}^{\n{F}}_i \right\} \ . 
\end{equation}
The measurement bins are uniformly spaced in time, so we denote the vectors of TOF and TOA bins by $\mb{t}^{\n{F}} \in \mbb{R}^{N_{\n{F}}}$ and $\mb{t}^{\n{A}} \in \mbb{R}^{N_{\n{A}}}$, respectively, such that
\begin{align*}
\mb{t}^{\n{A}}_j &= j \, \Delta_{\n{t}} + t_0 \\
\mb{t}^{\n{F}}_i &= (i-i_0) \, \Delta_{\n{t}} + t_0 \ ,
\end{align*}
where $\Delta_{\n{t}}$ is the TOF bin size, $t_0$ is the time of the first measured bin, and $i_0$ is a positive integer offset.
Importantly, the set of TOF bins is taken to be sufficiently large that it includes any neutron that could possibly contribute to a measurement.
This means that the set of TOA bins will be a contiguous subset of the set of TOF bins, and that $N_{\n{F}} > N_{\n{A}}$.
Equivalently, $R$ is a sparse matrix that is taller than it is wide.
Finally, we assume that the columns of $R$ sum to 1, so that $(\forall j) \sum_i R_{i,j} = 1$.
In \appx{ResolutionFunction} we show that this is because the uncertainty in TOA is modeled as a linear combination of temporal convolutions with the neutron pulse shape. 

\about{final model, unknowns}
Using this notation, the final equation for the expected measured neutron flux is given by
\begin{align}
\label{eq:forward_sample}
\bar{Y}_{\n{s}} &= \alpha_1 \left[ \varPhi \odot \left( \exp \left\{- Z D \right\} R \right) + \alpha_2 B \right] \, .
\end{align}
Note that both $D$ and $R$ are assumed to be known for the materials of interest and the particular neutron beamline. 
The unknowns that must be estimated are therefore $Z$, $\varPhi$, $B$, $\alpha_1$, and $\alpha_2$. It is clear that the dimension of the unknowns far exceeds twice the dimension of the measurements, $\bar{Y}_{\n{s}}$.
Consequently, the direct inversion of \eqref{eq:forward_sample} is very ill-posed.

\about{three assumptions}
In order to make the inversion problem well posed, we make three additional assumptions. 
First, assume that we have made an additional open beam measurement, $\bar{Y}_{\n{o}}$, with the sample removed.
Second, assume that the flux and background can be modeled with a low rank approximation.
And finally, assume there are two regions in our sample measurement, one in which the sample is not present, and another in which the sample has constant, non-zero areal density.

\about{(1) second measurement}
First, we will assume that a second open beam measurement is made with the sample removed.
In this case, $Z=\mb{0}$, and we have that
\begin{align}
\label{eq:forward_open}
\bar{Y}_{\n{o}} &= \varPhi + B \, ,
\end{align}
which uses the result \eq{1R1} from \appx{ResolutionFunction}.

\about{(2) reduction of dimensionality}
Second, in order to further improve the conditioning of the inversion problem, we will introduce the constraints that the neutron flux and background, $\varPhi$ and $B$, can be modeled with a rank-one approximation given by
\begin{align}
\label{eq:spectraPhi}
\varPhi &=  \mb{v}\mbs{\upvarphi}^\top           \\
\label{eq:spectraB}
B    &=  \mb{v} \mb{b}(\mbs{\uptheta})^\top \,,
\end{align}
where $\mb{v} \in \mbb{R}^{N_{\n{p}}}$, $\mbs{\upvarphi} \in \mbb{R}^{N_{\n{A}}}$, and
\begin{equation}
\mb{b}(\mbs{\uptheta})^\top = \exp \{ \mbs{\uptheta}^\top P \} \, ,
\label{eq:b-function}    
\end{equation}
where $\mbs{\uptheta} \in \mbb{R}^{N_{\n{b}}}$ is a low dimensional parameter vector,
and $P\in \mbb{R}^{N_{\n{b}} \times N_{\n{A}}}$ is a matrix of basis functions used to model the spectral characteristics of the background counts.
The specific form of $P$ is given in detail in Appendix~\ref{app:P-Matrix}.
In order to make the decomposition unique, we will also require that 
\begin{equation}
\mbb{1}^\top \mb{v} = N_{\n{p}} \,,
\label{eq:v-constraint}
\end{equation}
where $\mbb{1}$ denotes a column vector of $1$'s, i.e. $\mb{v}$ is normalized so that its components have an average value of $1$.

\about{(3) $\Omega_{\n{0}}$, $\Omega_{\n{z}}$ regions}
Third, we will assume that in the sample measurement field of view there are two regions of pixels, denoted by $\Omega_{\n{0}}$, the open beam region, and $\Omega_{\n{z}}$, the uniformly dense region.
In $\Omega_{\n{0}}$ no part of the sample is covering the beam.
In $\Omega_{\n{z}}$ the sample has uniform cross section.
More specifically, we assume
\begin{align}
\label{eq:Omega_z}
    (\forall i \in \Omega_{\n{z}}) \ Z_{i,*} &= \mb{z}^\top \\
\label{eq:Omega_0}
    (\forall i \in \Omega_{\n{0}}) \ Z_{i,*} &= \mbb{0}^\top \, ,
\end{align}
where $\mb{z} \in \mbb{R}^{N_{\n{m}}}$ is an unknown vector of areal densities and $\mbb{0}$ is a vector of zeros. These regions must be selected by the user based on knowledge of the sample properties, which is feasible in most applications. This can be achieved by leaving a part of the field of view unobscured to derive the $\Omega_{\n{0}}$ region. 
One possible method to pick the $\Omega_{\n{z}}$ region is to select a region in \textit{good faith}, perform a material decomposition and then check whether the resulting density, $Z$, is actually approximately uniform in $\Omega_{\n{z}}$.
If the sample does not readily provide a region of approximately equal areal density, a calibration sample could be added to the field of view to derive the $\Omega_{\n{z}}$ region.

\about{actual measurements}
Finally, the actual measurements $Y_{\n{s}}$ and $Y_{\n{o}}$ are assumed to be conditionally Poisson distributed, which is expressed as
\begin{align}
\label{eq:measurement_poisson}
Y_{\n{s}} \sim
\n{Poisson}( \bar{Y}_{\n{s}} )\ &| \ 
(Z,\, \mb{v},\, \mbs{\upvarphi},\, \mbs{\uptheta},\, \alpha_1,\, \alpha_2) \\
Y_{\n{o}} \sim \n{Poisson}( \bar{Y}_{\n{o}} ) \ &| \ 
(\mb{v},\, \mbs{\upvarphi},\, \mbs{\uptheta}) \nonumber \,.
\end{align}

\section{Estimation of Nuisance Parameters}
\label{sec:prep}

\about{goal}
In this section, we describe our procedure for estimating the nuisance parameters, $\mb{v}$, $\mbs{\upvarphi}$, $\mbs{\uptheta}$, $\alpha_1$, and $\alpha_2$.
This is a challenging problem because it is equivalent to estimation of the background, which tends to be large and spatially variable in neutron imaging.

\about{estimate $\mb{v}$}
We start by estimating $\mb{v}$ by observing that columns of the open beam scan, $\bar{Y}_{\n{o}}$, must all be proportional to $\mb{v}$, and that $\mbb{1}^\top \hat{\mb{v}}= N_{\n{p}}$. From this we obtain the estimate
\begin{equation}
\label{eq:v-estimation}
\hat{\mb{v}} = N_{\n{p}} \frac{ Y_{\n{o}} \mbb{1} }{ \mbb{1}^\top Y_{\n{o}} \mbb{1} } \, .
\end{equation}
Intuitively, this estimate simply averages the columns of $Y_{\n{o}}$ and scales the result so that \eq{v-constraint} holds.
While this estimator is not optimal for Poisson noise, we have found that it works well given the large number of pixels.


\about{define $\mb{q}(\mb{z})$:\\transmission in $\Omega_{\n{z}}$}
As stated in \eq{Omega_z} and \eq{Omega_0}, the measurement contains a uniformly dense region, $\Omega_{\n{z}}$, and a open beam region, $\Omega_{\n{0}}$. 
Then for each pixel in the region $\Omega_{\n{z}}$, the theoretical transmission spectrum through the sample is given by
\begin{equation}
\mb{q}(\mb{z})^\top = \exp \left\{ -\mb{z}^\top D \right\} R \, ,
\label{eq:q-function}
\end{equation}
where $\mb{q}(\mb{z}) \in \mbb{R}^{N_{\n{A}}}$ is the vector of transmission resulting from the material areal densities, $\mb{z}$. For each pixel in the region $\Omega_{\n{0}}$ we observe 100\% transmission in the open beam region, 
\begin{equation}
    \label{eq:q0-function}
    \mb{q}(\mbb{0})^\top = \mbb{1}^\top \ ,
\end{equation}
where we use the fact that $\mbb{1}^\top R = \mbb{1}^\top$, shown in \appx{ResolutionFunction}.

\about{define $\mbs{\upomega}_{\n{0}}$ and $\mbs{\upomega}_{\n{z}}$}
Let $\mbs{\upomega}_{\n{0}} \in \mbb{R}^{N_{\n{p}}}$ and $\mbs{\upomega}_{\n{z}} \in \mbb{R}^{N_{\n{p}}}$ be a weighting vectors that take the averages over pixels in those regions, $\Omega_{\n{0}}$ and $\Omega_{\n{z}}$, respectively. More precisely, let 
\begin{align*}
(\mbs{\upomega}_{\n{0}})_i
&=
\left\{
\begin{array}{ll}
\sfrac{1}{|\Omega_{\n{0}}|}  & \n{if} \, i\in \Omega_{\n{0}} \\
0     & \n{otherwise} 
\end{array}
\right. \\
(\mbs{\upomega}_{\n{z}})_i
&=
\left\{
\begin{array}{ll}
\sfrac{1}{|\Omega_{\n{z}}|}  & \n{if} \, i\in \Omega_{\n{z}} \\
0     & \n{otherwise} 
\end{array}
\right. \ .
\end{align*}
Then define the following three averages:
\begin{align}
\label{eq:y_o}
\bar{\mb{y}}_{\n{o}}^\top
&= \frac{\mbb{1}^\top \bar{Y}_{\n{o}}}{\mbb{1}^\top \mb{v}}
\\
\label{eq:y_sz}
\bar{\mb{y}}_{\n{sz}}^\top
&= \frac{\mbs{\upomega}_{\n{z}}^\top \bar{Y}_{\n{s}}}{\mbs{\upomega}_{\n{z}}^\top \mb{v}} \\
\label{eq:y_s0}
\bar{\mb{y}}_{\n{s0}}^\top
&= \frac{\mbs{\upomega}_{\n{0}}^\top \bar{Y}_{\n{s}}}{\mbs{\upomega}_{\n{0}}^\top \mb{v}} \, .
\end{align}
\about{derive term for $\bar{\mb{y}}_{\n{o}}$}
By using \eq{y_o}, \eq{forward_open}, \eq{spectraPhi}, and \eq{spectraB}, we then have that
\begin{align}
\label{eq:y_o_derivation}
\bar{\mb{y}}_{\n{o}}^\top &= \frac{\mbb{1}^\top \bar{Y}_{\n{o}}}{\mbb{1}^\top \mb{v}}
            = \frac{\mbb{1}^\top}{\mbb{1}^\top \mb{v}} \left( \varPhi + B \right) \nonumber\\
    &= \frac{\mbb{1}^\top \mb{v}}{\mbb{1}^\top \mb{v}}
       \left( \mbs{\upvarphi} + \mb{b}(\mbs{\uptheta}) \right)^\top 
            = \left( \mbs{\upvarphi} + \mb{b}(\mbs{\uptheta}) \right)^\top \, .
\end{align}
Rearranging terms then yields the following result
\begin{equation}
\mbs{\upvarphi} = \bar{\mb{y}}_{\n{o}} - \mb{b}(\mbs{\uptheta}) \, .
\label{eq:varphi}
\end{equation}
\about{derive term for $\bar{\mb{y}}_{\n{sz}}$}
Next, by using \eq{y_sz}, \eq{forward_sample}, \eq{spectraPhi}, \eq{spectraB}, and \eq{q-function} along with the expression for $\mbs{\upvarphi}$ from \eq{varphi}, we can derived the following expression for the average $\bar{\mb{y}}_{\n{sz}}^\top$ given by
\begin{align}
\label{eq:y_sz_derivation}
\bar{\mb{y}}_{\n{sz}}^\top &= \frac{\mbs{\upomega}_{\n{z}}^\top \bar{Y}_{\n{s}}}{\mbs{\upomega}_{\n{z}}^\top \mb{v}} \nonumber\\
    &= \frac{\alpha_1}{\mbs{\upomega}_{\n{z}}^\top \mb{v}} \mbs{\upomega}_{\n{z}}^\top 
    \left[ \varPhi \odot \left( \exp \left\{- \mbb{1} \mb{z}^\top D \right\} R \right) + \alpha_2 B \right] \nonumber\\
    &= \frac{\alpha_1}{\mbs{\upomega}_{\n{z}}^\top \mb{v}} \mbs{\upomega}_{\n{z}}^\top  \left[ ( \mb{v} \mbs{\upvarphi}^\top ) \odot \left( \mbb{1} \mb{q}(\mb{z})^\top \right) + \alpha_2 \mb{v} \mb{b}(\mbs{\uptheta})^\top \right] \nonumber\\
    &= \alpha_1 \left[ \mbs{\upvarphi} \odot \mb{q}(\mb{z}) + \alpha_2 \mb{b}(\mbs{\uptheta}) \right]^\top \nonumber\\
    &= \alpha_1 \left[ (\bar{\mb{y}}_{\n{o}} - \mb{b}(\mbs{\uptheta})) \odot \mb{q}(\mb{z}) + \alpha_2 \mb{b}(\mbs{\uptheta}) \right]^\top \ .
\end{align}

\about{derive term for $\bar{\mb{y}}_{\n{s0}}$}
Similar to \eq{y_sz_derivation}, an expression for $\bar{\mb{y}}_{\n{s0}}^\top$ can be found using \eq{y_s0} and \eq{q0-function},
\begin{align}
\label{eq:y_s0_derivation}
\bar{\mb{y}}_{\n{s0}}^\top &= \frac{\mbs{\upomega}_{\n{0}}^\top \bar{Y}_{\n{s}}}{\mbs{\upomega}_{\n{0}}^\top \mb{v}}
        = \alpha_1 \left[ (\bar{\mb{y}}_{\n{o}} - \mb{b}(\mbs{\uptheta})) \odot \mb{q}(\mb{z}) + \alpha_2 \mb{b}(\mbs{\uptheta}) \right]^\top \nonumber\\
    &= \alpha_1 \left[ \bar{\mb{y}}_{\n{o}} + (\alpha_2 - 1) \mb{b}(\mbs{\uptheta}) \right]^\top \, .
\end{align}

\about{optimization formulation}
By substituting in the measured neutron counts ($Y_{\n{s}}$, $Y_{\n{o}}$) for the expected measurements ($\bar{Y}_{\n{o}}$, $\bar{Y}_{\n{s}}$) in \eq{y_sz_derivation} and \eq{y_s0_derivation}, we can formulate a loss function. Then we can estimate the nuisance parameters by minimizing this loss function:
\begin{align}
\label{eq:NuisanceParamsEstimation}
(\hat{\mb{z}} , \hat{\alpha}_1, \hat{\alpha}_2, \hat{\mbs{\uptheta}})
=
\argmin_{\mb{z}, \alpha_1, \alpha_2, \mbs{\uptheta}}
\{
&\left\Vert 
\mb{y}_{\n{sz}} 
-
\mb{f}(\mb{z}, \alpha_1 , \alpha_2 , \mbs{\uptheta})
\right\Vert^2  
\nonumber\\
+ \beta & \left\Vert    
\mb{y}_{\n{s0}} 
-
\mb{f}(\mbb{0}, \alpha_1 , \alpha_2 , \mbs{\uptheta})
\right\Vert^2
\} \nonumber \\
\n{s. \, t.} \hspace{1cm}
        \mb{z} \geq \mbb{0} \hspace{0.4cm}& \nonumber \\
        \alpha_1, \alpha_2 \geq 0 \hspace{0.4cm}& \nonumber \\
        \mbs{\uptheta} \in \mbb{R}^{N_{\n{b}}} & \, ,
\end{align}
where the forward operators are defined by,
\begin{align}
\label{eq:f_operator}
\mb{f}(\mb{z}, \alpha_1 , \alpha_2 , \mbs{\uptheta} )
&= 
\alpha_1 \left[
\left( \mb{y}_{\n{o}} - \mb{b}(\mbs{\uptheta}) \right) \odot \mb{q}(\mb{z}) + \alpha_2 \mb{b}(\mbs{\uptheta})
\right]  \\
\mb{f}(\mbb{0}, \alpha_1 , \alpha_2 , \mbs{\uptheta} )
&= 
\alpha_1 \left[ \mb{y}_{\n{o}} + (\alpha_2 - 1) \mb{b}(\mbs{\uptheta})
\right]  \, ,
\end{align}
and the scalar $\beta$ is a weighting parameter that balances the associated costs between solving \eq{y_sz_derivation} and \eq{y_s0_derivation}. Since $\mbs{\upomega}_{\n{0}}^\top$ and $\mbs{\upomega}_{\n{z}}^\top$ perform averaging over many pixels, the Poisson distribution approximates a Gaussian via the Central Limit Theorem, which we use as a justification for the use of Least Squares estimators in \eq{NuisanceParamsEstimation}.

\about{final estimates}
This optimization yields the estimates $\hat{\mb{z}}$, $\hat{\alpha}_1$, $\hat{\alpha}_2$, and $\hat{\mbs{\uptheta}}$. The final parameter, $\mbs{\upvarphi}$, is then estimated via \eq{varphi} as
\begin{equation}
\hat{\mbs{\upvarphi}} = \mb{y}_{\n{o}} - \mb{b}(\hat{\mbs{\uptheta}}) \, ,
\label{eq:varphi-estimate}
\end{equation}
and the estimates for the flux and background
\begin{align}
    \label{eq:Phihat_Bhat}
    \hat{\varPhi} &= \hat{\mb{v}} \hat{\mbs{\upvarphi}}^\top \\
    \hat{B} &= \hat{\mb{v}} \mb{b}(\hat{\mbs{\uptheta}})^\top \nonumber \, .
\end{align}

\section{Areal Density Reconstruction}

\about{goal}
In this section, we describe how we reconstruct the isotope areal density, $Z$, based on a statistical model of the neutron counting statistics. The estimated nuisance parameters from \sctn{prep} are considered known at this point.

\about{statistical model,\\ML estimate}
We assume that the actual neutron sample measurements, $Y_{\n{s}}$, are independent Poisson distributed with conditional mean $\bar{Y}_{\n{s}}$.
More precisely, if we define the operator
\begin{align}
F(Z) =
\hat{\alpha}_1 \left[ \hat{\varPhi} \odot \left( \exp \left\{- Z D \right\} R \right) + \hat{\alpha}_2 \hat{B} \right] \, ,
\label{eq:F_operator}
\end{align}
then we assume that 
$$
Y_{\n{s}} \sim \n{Poisson}( F(Z) ) \ | \ Z \, ,
$$
and the conditional mean of the observations given the unknown $Z$ is therefore $$\mbb{E}\left[ Y_{\n{s}} | Z \right] = F(Z) \, .$$
For a Poisson random matrix $Y$ with mean $\varLambda$, the negative log-likelihood is
\begin{align*}
    \mc{L}( Y | \varLambda ) &= - \log f_{\n{Y|\Lambda}}(Y|\varLambda )   \\
         &= \sum_{i,j} \varLambda_{i,j} - Y_{i,j} \log  \varLambda_{i,j} + \log(Y_{i,j}!) \, .
\end{align*}
Then our maximum likelihood estimate for the material areal density is given by
\begin{equation}
\label{eq:MaterialDecompositionEstimation}
\hat{Z} = \argmin_{Z \in \mbb{R}^{N_{\n{p}} \times N_{\n{m}}}} \mc{L}( Y_{\n{s}} | F(Z) ) \, .
\end{equation}

\section{Volumetric Density Reconstruction}
\label{sec:tomography}

In this section we describe a 3D tomographic extension to the 2D case that is possible when measurements of multiple rotational views are available. This will be a two step approach involving first reconstructing the areal densities, $Z$, for each view, and then performing tomographic reconstruction of the volumetric densities, $X$.

\about{reconstruct $Z$}
Let $\{ Y_{\n{s}}^{(k)} \}_{k = 1}^{N_{\n{r}}}$ be the set of $N_{\n{r}}$ measurements at distinct views. Corresponding to these measurements, we will reconstruct $N_{\n{r}}$ areal density views $\{ Z^{(k)} \}_{k = 1}^{N_{\n{r}}}$ using the material decomposition of \eq{MaterialDecompositionEstimation}. We assume all nuisance parameters except $\alpha_1$ are independent of view index and thus only need to be computed once from a single view. The $\{ \alpha_1^{(k)} \}_{k = 1}^{N_{\n{r}}}$ set of parameters are assumed to be dependent on the view index, $k$, and are estimated using the closed form expression\footnote{This expression is a corollary of \eq{y_s0_derivation} and its derivation is omitted.}
\begin{align}
\hat{\alpha}_1^{(k)}
=
\frac
{\left[ \mb{y}_{\n{s0}}^{(k)} \right]^\top \mbb{1}}
{\left[
\mb{y}_{\n{o}}
+ (\hat{\alpha}_2 - 1) \mb{b}(\hat{\mbs{\uptheta}})
\right]^\top \mbb{1}} \, ,
\end{align}
where
$
[ \mb{y}_{\n{s0}}^{(k)} ]^\top = \sfrac{([\mbs{\upomega}_{\n{0}}^{(k)}]^\top Y_{\n{s}}^{(k)})}{([\mbs{\upomega}_{\n{0}}^{(k)}]^\top \hat{\mb{v}})}
$
correspond to the user selected open beam regions, $\Omega_{\n{0}}^{(k)}$, while $\Omega_{\n{z}}^{(k)}$ regions need not necessarily be available for every rotational view.

\about{decomposition $Z=AX$}
The estimated set of areal densities is stacked, and we assume the following decomposition
\begin{align}
    \hat{Z} =
    \begin{bmatrix}
        \hat{Z}^{(1)} \\
        \vdots \\
        \hat{Z}^{(N_{\n{r}})}
    \end{bmatrix}
    = AX \, ,
\label{eq:Z_AX_decomposition}
\end{align}
where the areal density estimates, $\hat{Z} \in \mbb{R}^{(N_{\n{r}}N_{\n{p}}) \times N_{\n{r}}}$, are in units of $\n{mol}/\n{cm}^2$, the tomographic projector, $A \in \mbb{R}^{(N_{\n{r}}N_{\n{p}}) \times N_{\n{v}}}$, is in units of $\n{cm}$, and the volumetric densities, $X \in \mbb{R}^{N_{\n{v}} \times N_{\n{m}}}$ are in units of $\n{mol}/\n{cm}^3$ and comprised of $N_{\n{v}}$ voxels. Then $X_{j,m}$ represents the volumetric density of the $j^{\n{th}}$ voxel for the $m^{\n{th}}$ isotope in the sample, and $A_{i,j}$ integrates along the tomographic projection line of the $j^{\n{th}}$ voxel and the $i^{\n{th}}$ measurement projection.

\about{volumetric density estimates}
Finally, computing the volumetric densities, $X$, boils down to solving \eq{Z_AX_decomposition} which can be decomposed by columns:
\begin{align*}
    \begin{bmatrix}
        | & & |\\
        \hat{Z}_{*,1} & \hdots & \hat{Z}_{*,N_{\n{m}}}\\
        | & & |
    \end{bmatrix}
    =
    A
    \begin{bmatrix}
        | & & |\\
        X_{*,1} & \hdots & X_{*,N_{\n{m}}}\\
        | & & |
    \end{bmatrix} \, .
\end{align*}
Consequently for each isotope the volumetric density $X_{*,m}$ is reconstructed from $\hat{Z}_{*,m}$ as the solution of
\begin{equation}
    \label{eq:ct}
    \hat{X}_{*,m} = \argmin_{\mbb{0} \leq \mb{x} \in \mbb{R}^{N_{\n{p}}} } \left\{ \Vert A \mb{x} - \hat{Z}_{*,m} \Vert^2 + r_m(\mb{x}) \right\} \, ,
\end{equation}
using a tomographic reconstruction algorithm, where $r_m(\cdot)$ is a 3D regularization term of the $m^{\n{th}}$ isotope. Although \eq{ct} specifies the use of MBIR reconstruction, other tomographic reconstruction methods can be used. However, we note that neutron tomography typically must use sparse views due to the slow view acquisition times, and that traditional CT reconstruction methods, such as filtered back projection, result in severe artifacts for the sparse view case. So we believe that MBIR reconstruction is a good choice for this application.

\section{Optimization Methods}
\label{sec:optimization}

The key steps in the reconstruction of \eq{NuisanceParamsEstimation} and \eq{MaterialDecompositionEstimation} require the solution of non-convex optimization problems.
In this section, we discuss the techniques we use to implement robust strategies for computing their solutions.

\subsection{Preconditioning}

A practical issue in optimization is that both \eq{NuisanceParamsEstimation} and \eq{MaterialDecompositionEstimation} incorporate loss functions with exponential terms such as $\mb{b}(\mbs{\uptheta})^\top = \exp \{ \mbs{\uptheta}^\top P \}$ and $\mb{q}(\mb{z})^\top = \exp \left\{ -\mb{z}^\top D \right\} R$. These terms can have large dynamic range and can therefore be sensitive to small changes in their arguments. 

In order to limit the dynamic range of these exponential terms, we defined $P$ in \eq{DefinitionOfP} so that its rows have unit norm.
We can do the same for the dictionary matrix $D$ by defining
\begin{equation}
\tilde{D} = C ^{-1} D \ , 
\end{equation}
where $C$ is a diagonal matrix such that $C_{i,i}= \Vert D_{i,*} \Vert$.
However, this also requires that we define the new functions
\begin{align}
\tilde{\mb{q}}( \tilde{\mb{z}})^\top 
    &= \exp \left\{ -\tilde{\mb{z}}^\top \tilde{D} \right\} R \\
\tilde{F}(\tilde{Z}) &=
\hat{\alpha}_1 \left[ \hat{\varPhi} \odot \left( \exp \left\{- \tilde{Z} \tilde{D} \right\} R \right) + \hat{\alpha}_2 \hat{B} \right] \, ,
\end{align}
that are used to replace the functions of \eq{q-function} and \eq{F_operator}.

After the modified optimization problems are solved, the solution of the original problem can be found using
\begin{align}
\mb{z}^\top &= \tilde{\mb{z}}^\top C \\
Z &= \tilde{Z} C  \, .
\end{align}

\subsection{Initialization}
\label{sec:initialization}

Since the optimization problems are not convex, it is important to select good initializations that lead to solutions that are close to the global minimum.
We first focus on the initialization of the optimization in \eq{NuisanceParamsEstimation} of the nuisance parameter estimation. The initial values that we use are
\begin{align}
\alpha_2 &= 1 \\
\label{eq:init_alpha_1}
\alpha_1 &=  \frac
{\mbs{\upomega}_{\n{0}}^\top Y_{\n{s}}}
{\mbs{\upomega}_{\n{0}}^\top Y_{\n{o}}}\\
\label{eq:init_theta}
\mbs{\uptheta}^\top &= 
\log \left( \min_i \left\lbrace
\frac
{(\mb{y}_{\n{sz}})_i}
{(\mb{y}_{\n{o}})_i}
\right\rbrace
\frac{\mb{y}_{\n{o}}}{\alpha_1 \alpha_2}
\right)^\top P^\dagger \\
\label{eq:init_z}
\mb{z}^\top &= - \log(\mb{q}^\top) (D R)^\dagger \, ,
\end{align}
where $(\cdot)^\dagger$ denotes the Moore–Penrose pseudoinverse and
$$
\mb{q} =
\left|
\frac
{\mb{y}_{\n{s}}/\alpha_1 - \alpha_2 \mb{b}(\mbs{\uptheta})}
{\mb{y}_{\n{o}} - \mb{b}(\mbs{\uptheta})}
\right|
\,
$$
is the average measured transmission, given the initial nuisance parameters, $\alpha_1$, $\alpha_2$, and $\mbs{\uptheta}$. This initialization is motivated by \eq{y_s0_derivation} and the following crude approximations:
\begin{align*}
\hat{\alpha}_2
&\approx 1 \\
\mb{y}_{\n{s0}} 
&\approx \hat{\alpha}_1 \left[ \mb{y}_{\n{o}} + (\hat{\alpha}_2 - 1) \mb{b}(\hat{\mbs{\uptheta}}) \right] \\
\hat{\alpha}_1 \hat{\alpha}_2 \mb{b}(\hat{\mbs{\uptheta}})
&\approx \min_i \left\lbrace
\frac
{(\mb{y}_{\n{sz}})_i}
{(\mb{y}_{\n{o}})_i}
\right\rbrace
\mb{y}_{\n{o}} \\
\exp\{-ZD\}R &\approx \exp\{-ZDR\} \, .
\end{align*}
For the optimization of \eq{MaterialDecompositionEstimation}, we initialize
\begin{equation}
Z = - \log(Q) (D R)^\dagger \, ,
\label{eq:init_Z}
\end{equation}
where the measured transmission
$$
Q = 
\left|
\frac
{Y_{\n{s}}/\hat{\alpha}_1 - \hat{\alpha}_2 \hat{B}}
{Y_{\n{o}} - \hat{B}}
\right| \,
$$
is computed given the final nuisance parameters.

\section{Experimental Results}
\label{sec:Results}

In this section we present results of the TRINIDI algorithm applied to both simulated and experimentally measured data. \sctn{Methods} outlines the experimental settings and notes about the evaluation of the results.
In \sctn{2DSimulated}, we demonstrate the method using 2D simulated radiographs. In \sctn{2DExperimental}, we provide similar 2D results for experimentally measured data using a known phantom. 
Finally in \sctn{3DExperimental}, we present results for fully tomographic 3D reconstruction of nuclear fuel pellets that are verified using independent mass spectroscopy measurements.

\subsection{Methods}
\label{sec:Methods}

\about{beamline}
Experiments in this work were performed at the Flight-Path-5 beamline at LANSCE~\cite{lisowski2006alamos}. In this facility, high energy spallation neutrons are slowed down by a high intensity ambient temperature water moderator resulting in a $20 \, \n{Hz}$ pulsed neutron beam with a wide energy spectrum. A detailed description of this facility including a characterization of the incident neutron spectrum can be found in~\cite{MOCKO2008455}.

\about{specs}
The neutrons travel along a flight path of length of $L = 10.4 \, \n{m}$ through the sample onto a TOF imaging detector. The neutron sensitive multi-channel plate detector works in conjunction with four Timepix readout chips, described in more detail in~\cite{tremsin2009detection}. The detector has $N_{\n{p}} = 512 \!\times\! 512$ pixels with a $55 \!\times\! 55 \, (\n{\upmu m})^2$ pixel pitch i.e. covering a field of view of approximately $28 \!\times\! 28 \, (\n{mm})^2$, and frames are recorded at a rate of, $\sfrac{1}{\Delta_{\n{t}}} \approx \sfrac{1}{30 \n{ns}}\approx 33.3 \, \n{MHz}$.

\about{resolution function}
The resolution function is modeled as an energy dependent weighted sum of two chi-squared probability density functions based on the work in~\cite{lynn2002neutron}. As described in \appx{ResolutionFunction}, the energy dependent pulse shape is approximated by a weighted sum of convolutions of $K$ kernels. We choose $K=5$ since we found that this gave sufficiently accurate representation of the resolution function.

\about{Define Estimated Open Beam\\ and Estimated Background}
In order to assess the accuracy of the nuisance parameter estimation of \eq{NuisanceParamsEstimation}, we plot $\mb{y}_{\n{s0}}$ (the measurement in the open beam region) and the fit of $\mb{y}_{\n{s0}}$ (the effective open beam) given by
\begin{equation}
\label{eq:Omega0_fit}
\mb{f}(\mbb{0}, \alpha_1 , \alpha_2 , \mbs{\uptheta})
=
\alpha_1 \left[
\mb{y}_{\n{o}} + (\alpha_2 - 1) \mb{b}(\mbs{\uptheta})
\right]\ \, ,
\end{equation}
where \eq{Omega0_fit} is the expected measurement when there is $100 \, \%$ transmission (i.e. zero material).
We also plot $\mb{y}_{\n{sz}}$ (the measurement in the uniformly dense region) and the fit of $\mb{y}_{\n{sz}}$ given by 
\begin{align}
\label{eq:Omegaz_fit}
\mb{f}(\mb{z}, \alpha_1 , \alpha_2 , \mbs{\uptheta} )
=
\alpha_1 \left[
\left( \mb{y}_{\n{o}} - \mb{b}(\mbs{\uptheta}) \right) \odot \mb{q}(\mb{z}) + \alpha_2 \mb{b}(\mbs{\uptheta})
\right] \, ,
\end{align}
where \eq{Omegaz_fit} is the expected measurement when there is a uniform material density, $\mb{z}$, attenuating the beam.
Lastly, we plot the effective background given by
\begin{align}
\label{eq:effective_background}
\mb{f}(\mbs{\infty}, \alpha_1 , \alpha_2 , \mbs{\uptheta} )
= \alpha_1 \alpha_2 \mb{b}(\mbs{\uptheta}) \, ,
\end{align}
where \eq{effective_background} is the expected measurement when there is there is $0 \, \%$ transmission (i.e. infinite material).

\about{normalization of results}
We display normalized density images for more accessible interpretation of our density estimates. The scales of the densities differ widely from one another so that instead of directly displaying the all $Z_{*,m}$ with a different color map, we choose to display all $Z_{*,m} / \mb{z}_m$ with the same color map. The scalars $\mb{z}_m$ are the ground truth average density of the $m^{\n{th}}$ isotope (or some other adequate normalization). Thus, the normalized densities, $Z \diag(\mb{z})^{-1}$, are unitless, where $1$ corresponds to the values of $\mb{z}$ i.e. ground truth. A similar treatment is applied to the volumetric densities, $X$.


\about{Grab-bag}
Finally, for the cross section dictionary, $D$, we use tabulated data, of the neutron total cross section at a temperature of $293.6 \ \n{K}$ from ENDF/B-VIII.0\cite{conlin2018release}.

We compare our reconstruction results to the reconstruction using the linear model from~\cite{balke2021hyperspectral} which we refer to as linear baseline reconstruction or $Z^{\n{lin}}$. In an attempt to further compare our results to existing methods, we tried processing the transmission spectra from our experiments using SAMMY~\cite{Larson2008}. However, due to the significant noise, we were unable to get the software to function properly.

All tomographic reconstructions were performed using the super-voxel model-based iterative reconstruction (SVMBIR) software package~\cite{svmbir-2020}, with parameters selected to maximize subjective image quality,
and all optimizations were performed using the accelerated proximal gradient method (APGM) with robust line search\footnote{This algorithm was selected due to the convenient availability of a reliable implementation rather than any expectation of optimum suitability for this problem, and it is expected that other optimization algorithms may provide faster convergence. In particular, after completion of the computational experiments reported here, we observed that the Broyden–Fletcher–Goldfarb–Shannon (BFGS) algorithm appears to converge significantly faster than APGM.} as part of the scientific computational imaging code (SCICO) software package~\cite{Balke2022}. The $\beta$ parameter from \eq{NuisanceParamsEstimation} is chosen to be equal to $1$ in \sctn{2DSimulated} and \sctn{3DExperimental} and chosen to be equal to $0$ in \sctn{2DExperimental} because in this experiment, there is no $\Omega_{\n{0}}$ region available. The number of background basis functions is chosen to be $N_{\n{b}} = 5$.

\subsection{2D Reconstruction from Simulated Data}
\label{sec:2DSimulated}

\begin{figure}[ht]
\vspace{-0.09cm} 
\centering
{\adjincludegraphics[width=1.0\columnwidth,
trim={{0.135\width} {0.05\height} {0.175\width} {0.0\height}}
,clip]{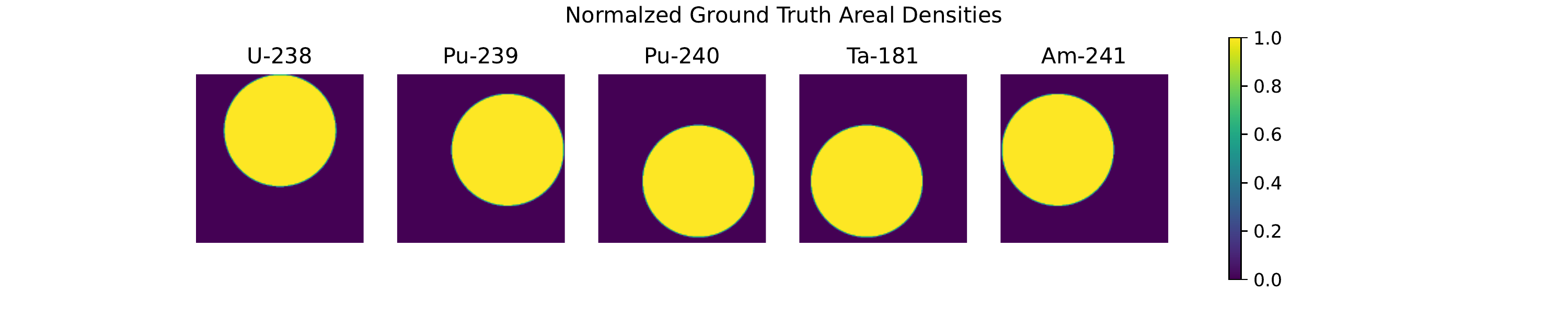}}
\caption[]{Normalized ground truth areal densities ($Z \diag(\mb{z})^{-1}$, [unitless]) used in the simulated data set of \sctn{2DSimulated}.}
\label{fig:mcsim_phantom}
\vspace{-0.09cm} 
\end{figure}

\about{goal/phantom}
In this experiment, we reconstruct a 2D phantom from Monte-Carlo simulated data.
\fig{mcsim_phantom} illustrates the structure of the simulated 2D phantom which is composed of a series of overlapping disks each formed by a distinct material isotope.
The other parameters of the simulation were chosen to approximately mimic a typical experimental scenario.

\begin{figure}[htb]
\vspace{-0.09cm} 
\centering
{\adjincludegraphics[width=1.0\columnwidth,
trim={{0.05\width} {0.04\height} {0.06\width} {0.11\height}}
,clip]{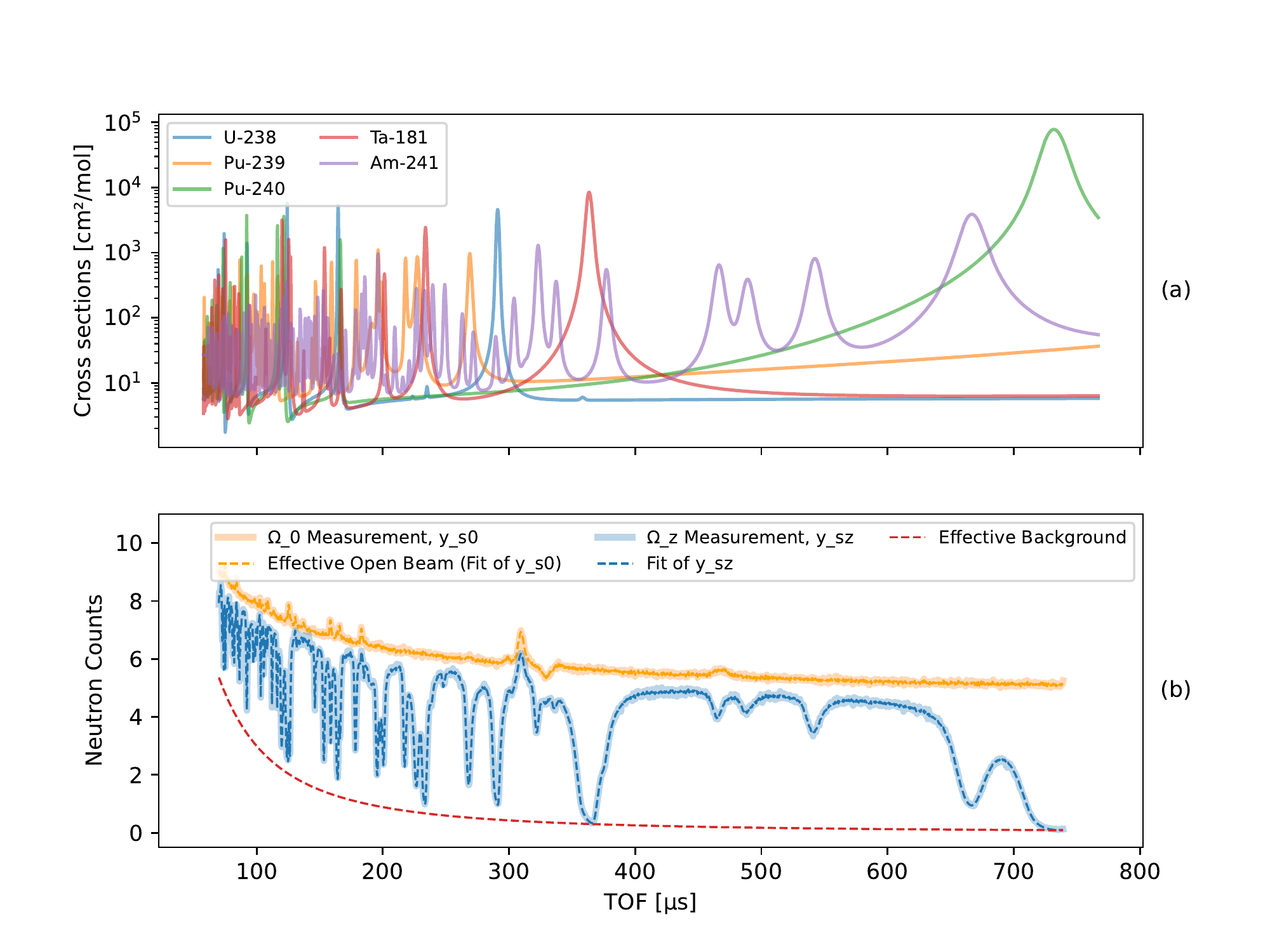}}
\caption[]{Nuisance parameter estimation fits of the simulated data set.\\
(a) Cross section dictionary, $D$, with $^{238}\n{U}$, $^{239}\n{Pu}$, $^{240}\n{Pu}$, $^{181}\n{Ta}$, and $^{241}\n{Am}$ isotopes in units of $\n{cm^2/mol}$.\\
(b) $\Omega_{\n{0}}$ measurement $\mb{y}_{\n{s0}}$ (yellow), effective open beam (yellow dashed), $\Omega_{\n{z}}$ measurement $\mb{y}_{\n{sz}}$ (blue), fit of $\mb{y}_{\n{sz}}$ (blue dashed), effective background (red dashed).}
\label{fig:mcsim_prep_fit}
\vspace{-0.09cm} 
\end{figure}

The ground truth areal densities were chosen to represent $N_{\n{m}}=5$ distinct isotopes, where each areal density map, $Z_{*,m}$, consists of a disk of constant density $\mb{z}_m$ with a surrounding area of zero material. Each map has $N_{\n{p}} = 128  \!\times\!  128$ pixels. The numerical values of the densities, $\mb{z}$, can be found in \tab{simulated_data} and were chosen to yield a visible distribution of resonances in the measurement spectra. The isotopes used are $^{238}\n{U}$, $^{239}\n{Pu}$, $^{240}\n{Pu}$, $^{181}\n{Ta}$, $^{241}\n{Am}$ and their respective cross sections are shown in \fig{mcsim_prep_fit}(a).

\about{generate measurements}
The simulated measurement counts, $Y_{\n{s}}$ and $Y_{\n{o}}$, were generated by first computing their sample means using \eq{forward_sample} and \eq{forward_open}, and then generating pseudo-random Poisson samples corresponding to those means.
In order to make the simulation representative of a typical physical experiment, the values for $\mb{v}$, $\mbs{\uptheta}$, $\alpha_1$, $\alpha_2$, $\mbs{\upvarphi}$, and the TOF sampling interval are chosen to be the same as the estimates from the real data covered in \sctn{2DExperimental}. 

\begin{figure}[ht]
\vspace{-0.09cm} 
\centering
{\adjincludegraphics[width=1.0\columnwidth,
trim={{0.15\width} {0.06\height} {0.15\width} {0.02\height}}
,clip]{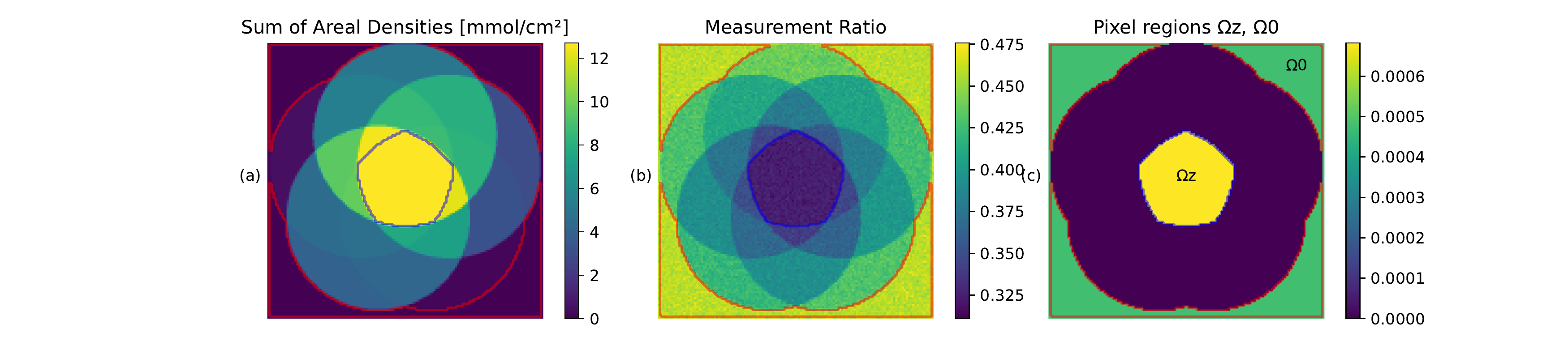}}
\caption[]{Open beam region, $\Omega_{\n{0}}$ (red line), and uniformly dense region, $\Omega_{\n{z}}$ (blue line), in simulated data set. \\
(a) Sum of ground truth areal densities ($Z \mbb{1}$). \\
(b) TOF integrated measurement ratio, $\sfrac{Y_{\n{s}} \mbb{1}}{Y_{\n{o}} \mbb{1}}$. \\
(c) Pixel regions: $\Omega_{\n{0}}$ green, $\Omega_{\n{z}}$ yellow.}
\label{fig:mcsim_regions}
\vspace{-0.09cm} 
\end{figure}

\about{$\Omega_{\n{z}}$ and $\Omega_{\n{0}}$}
\fig{mcsim_regions} shows the summed areal densities over all isotopes, the TOF integrated measurement ratio $\sfrac{Y_{\n{s}} \mbb{1}}{Y_{\n{o}} \mbb{1}}$, and the regions $\Omega_{\n{z}}$ and $\Omega_{\n{0}}$ used for this experiment. 
The set $\Omega_{\n{z}}$ was taken to be the region in the center of the phantom in which all the disks overlap, and the set $\Omega_{\n{0}}$ was taken to be the region surrounding the disks that is not covered by any material.

\begin{table}[ht]
\vspace{-0.09cm} 
\centering
\caption{\label{tab:simulated_data} Ground truth and estimated values of nuisance parameters for the simulated data sample.}
\scalebox{0.92}{
    \begin{tabular}{@{}c|c@{}}
    \toprule
    Quantity & Ground Truth\\ 
    \midrule
    $\left[ \begin{matrix} \n{isotopes} \\ \mb{z}^\top \, ^{[\n{mmol/cm^2}]} \end{matrix} \right]$
    & $\left[ \begin{matrix} ^{238}\n{U} & ^{239}\n{Pu} & ^{240}\n{Pu} & ^{181}\n{Ta} & ^{241}\n{Am} \\ 5.00 & 3.00 & 0.200 & 4.00 & 0.500 \end{matrix}\right]$\\
    $\alpha_1, \ \ \alpha_2$ &  $0.483, \ \ 0.685$\\
    $\mbs{\uptheta}^\top$ &  $\left[\begin{matrix}29.9 & -56.1 & 5.39\end{matrix}\right]$\\
    \bottomrule
    \toprule
    Quantity & Estimates\\ 
    \midrule
    $\left[ \begin{matrix} \n{isotopes} \\ \hat{\mb{z}}^\top \, ^{[\n{mmol/cm^2}]} \end{matrix} \right]$
    & $\left[\begin{matrix} ^{238}\n{U} & ^{239}\n{Pu} & ^{240}\n{Pu} & ^{181}\n{Ta} & ^{241}\n{Am} \\ 5.00 & 2.99 & 0.194 & 3.93 & 0.493 \end{matrix} \right]$ \\
    $\hat{\alpha}_1, \ \ \hat{\alpha}_2$ &  $0.481, \ \ 0.687$\\
    $\hat{\mbs{\uptheta}}^\top$ &  $\left[\begin{matrix}27.7 & -59.6 & 3.50\end{matrix}\right]$\\
    \bottomrule
    \end{tabular}
}
\vspace{-0.09cm} 
\end{table}

\about{resulting nuisance parameters}
\tab{simulated_data} lists the results of the nuisance parameter estimation using \eq{NuisanceParamsEstimation}. 
Notice that the parameters of interest, $\hat{\mb{z}}$, are accurately estimated within approximately $3\%$.
The estimation of the nuisance parameters, $\alpha_1$, $\alpha_2$ and $\mbs{\uptheta}$, is more variable; however, we note that accurate estimation of nuisance parameters may not always be necessary as long as the effective open beam and background are accurately modeled.

\about{fits}
\fig{mcsim_prep_fit}(b) shows plots corresponding to the fitted curves of the nuisance parameter estimation. The orange and blue curves are the average sample measurements, $\mb{y}_{\n{s0}}$ and $\mb{y}_{\n{sz}}$, and the dashed lines of the same respective colors are their corresponding fits. Note that the measurements and fits visually align very well. The red dashed line is the resulting effective background. 

\begin{figure}[htb]
\vspace{-0.09cm} 
\centering
{\adjincludegraphics[width=1.0\columnwidth,
trim={{0.06\width} {0.0\height} {0.1\width} {0.11\height}}
,clip]{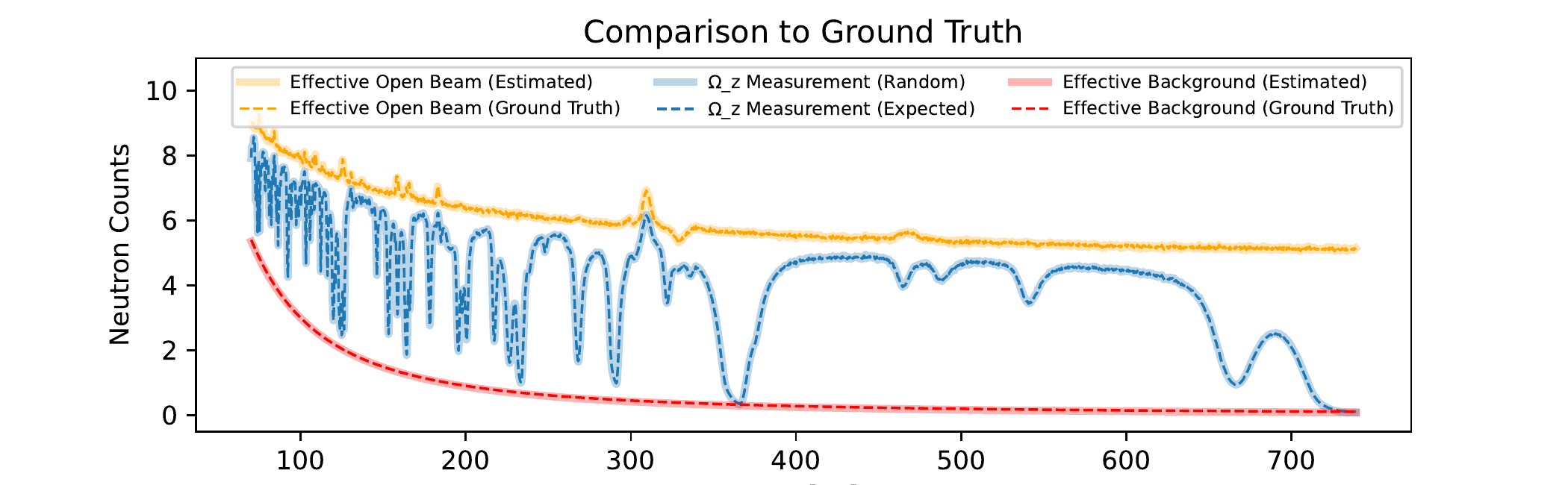}}
\caption[]{Comparison of estimated vs. ground truth spectra resulting from the nuisance parameter estimation of the simulated data set.}
\label{fig:mcsim_prep_GT_comparison}
\vspace{-0.09cm} 
\end{figure}

\about{direct comparison to ground truth}
\fig{mcsim_prep_GT_comparison} shows a direct comparison to the ground truth spectra. The effective open beam (yellow lines) and background (red lines) are compared directly their respective ground truth spectra (dashed lines). 
Notice that all three fits are quite accurate, even though the underlying nuisance parameter estimates in \tab{simulated_data} where not as accurate.

\begin{figure}[ht]
\vspace{-0.09cm} 
\centering
{\adjincludegraphics[width=0.60\columnwidth,
trim={{0.07\width} {0.46\height} {0.09\width} {0.11\height}}
,clip]{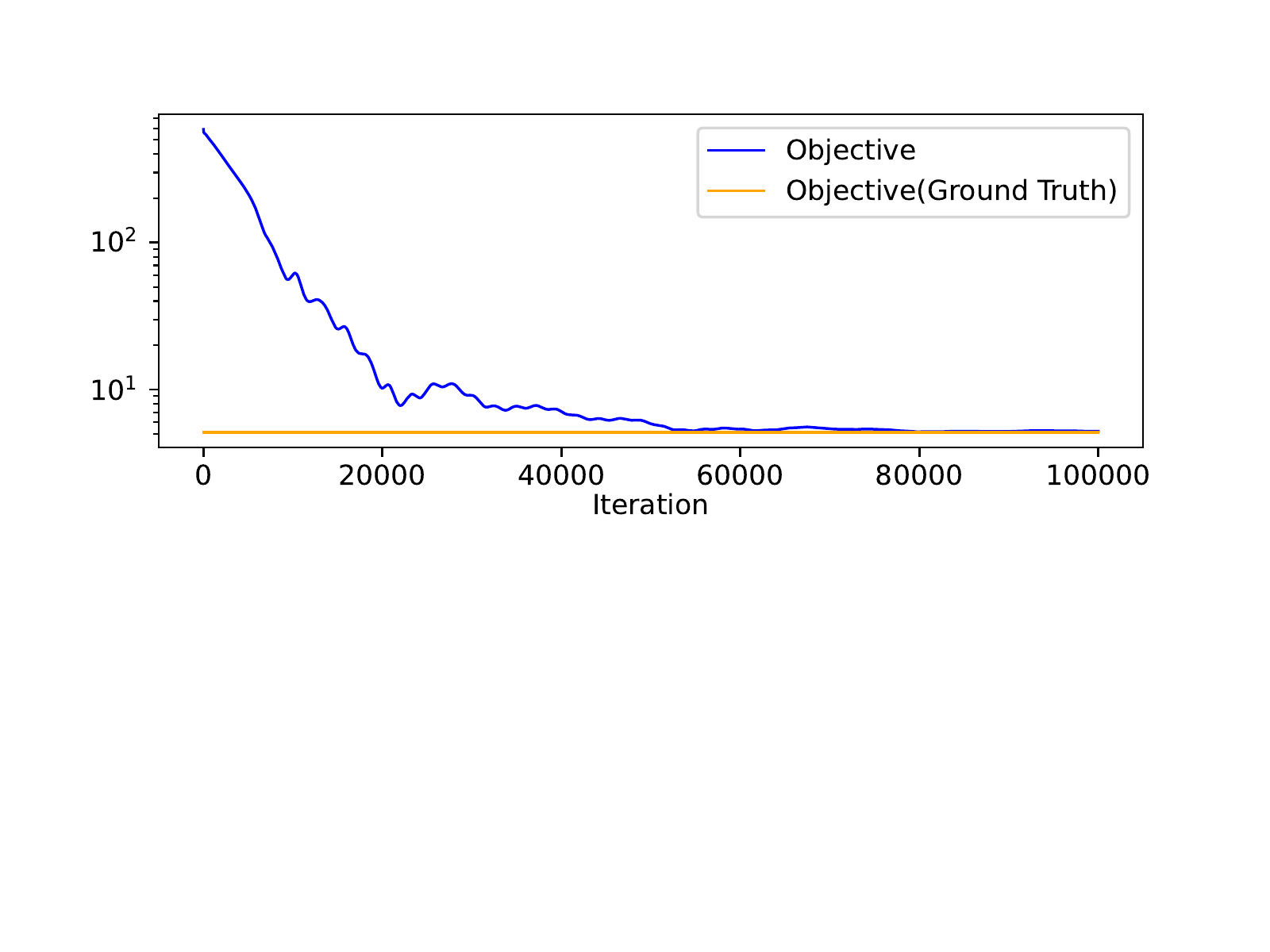}}
\caption[]{Convergence of nuisance parameter estimation for simulated data.}
\label{fig:mcsim_convergence}
\vspace{-0.09cm} 
\end{figure}

\about{convergence}
\fig{mcsim_convergence} shows the objective of the optimization as a function of iteration number. Although the optimization takes many iterations and is not monotone, the objective eventually converges to approximately the same value as the ground truth.


\begin{figure*}[ht]
\vspace{-0.09cm} 
\centering
{\adjincludegraphics[width=1.6\columnwidth,
trim={{0.08\width} {0.08\height} {0.17\width} {0.02\height}}
,clip]{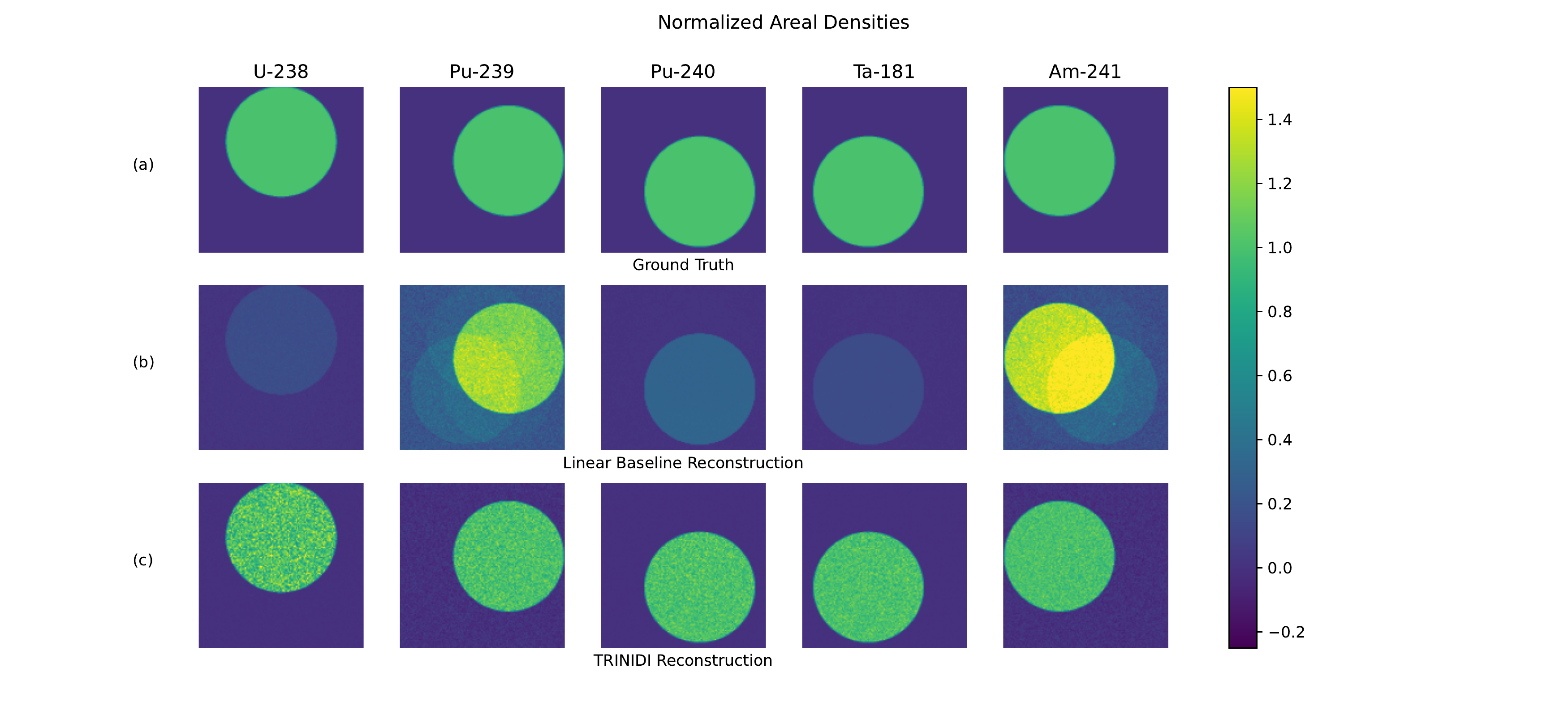}}
\caption[]{Normalized areal densities results for the simulated data [unitless].\\
(a) Normalized ground truth areal density ($Z \diag(\mb{z})^{-1}$).\\
(b) Normalized linear baseline~\cite{balke2021hyperspectral} reconstructed areal density ($Z^{\n{lin}} \diag(\mb{z})^{-1}$),\\
(c) Normalized TRINIDI reconstructed areal density ($\hat{Z} \diag(\mb{z})^{-1}$).}
\label{fig:mcsim_Z--Z0--Z_hat}
\vspace{-0.09cm} 
\end{figure*}

\about{2D results}
\fig{mcsim_Z--Z0--Z_hat}(a) shows the normalized ground truth areal densities, \fig{mcsim_Z--Z0--Z_hat}(b) shows the linear baseline~\cite{balke2021hyperspectral} reconstructed areal density, and \fig{mcsim_Z--Z0--Z_hat}(c) shows the corresponding TRINIDI reconstructed areal density.
The linear baseline reconstruction deviates significantly from the ground truth.
Notice that using TRINIDI each isotope map is accurately reconstructed from the simulated data. Also, even though the disks significantly overlap with one another, especially in the center of the field of view, there are no visible artifacts, which is not the case for the linear baseline.

It is worth pointing out that the cross section dictionary, $D$, contains all five isotopes, however, most pixels in the ground truth do not have all five isotopes present. For example, in the very top-center region only $^{238}\n{U}$ is present. The TRINIDI reconstructions show that $^{238}\n{U} $ is reconstructed accurately as non-zeros while the other isotopes are reconstructed accurately as (close to) zeros. This indicates that overpopulation of the cross section dictionary will likely not result in degradation of the reconstruction, however, we suggest that unnecessary overpopulation should be avoided.


\begin{table}[ht]
\vspace{-0.09cm} 
\centering
\caption{\label{tab:simulated_data2} Mean and standard deviation of areal density maps of ground truth ($Z$), linear baseline reconstruction~\cite{balke2021hyperspectral} ($Z^{\n{lin}}$), and TRINIDI reconstruction ($\hat{Z}$) in units of $[\n{mmol/cm^2}]$ of the simulated data sample.}
\scalebox{0.92}{
    \begin{tabular}{@{}c|ccc@{}}
    \toprule
    Isotopes & Ground Truth & Linear Baseline~\cite{balke2021hyperspectral} & TRINIDI \\ 
    \midrule
    $^{238}\n{U} $ & $5.0 \pm 0.0$ & $0.825 \pm 0.082$ & $\mathbf{5.065 \pm 1.045}$\\
    $^{239}\n{Pu}$ & $3.0 \pm 0.0$ & $3.613 \pm 0.336$ & $\mathbf{3.009 \pm 0.341}$\\
    $^{240}\n{Pu}$ & $0.2 \pm 0.0$ & $0.062 \pm 0.001$ & $\mathbf{0.202 \pm 0.021}$\\
    $^{181}\n{Ta}$ & $4.0 \pm 0.0$ & $0.631 \pm 0.021$ & $\mathbf{4.015 \pm 0.376}$\\
    $^{241}\n{Am}$ & $0.5 \pm 0.0$ & $0.703 \pm 0.067$ & $\mathbf{0.499 \pm 0.039}$\\
    \bottomrule
    \end{tabular}
}
\vspace{-0.09cm} 
\end{table}

\about{quantitative results}
\tab{simulated_data2} shows the mean and standard deviation of the density estimates, $\hat{Z}_{*,m}$ for each of the disks of material individually compared to the ground truth ($Z$) and the linear baseline ($Z^{\n{lin}}$). Note that even though there is significant noise, the mean estimates using TRINIDI are very close to the ground truth densities. Also note that the different isotopes have different estimation accuracy. In comparison, the linear baseline estimates are not quantitatively accurate.

\begin{figure}[htb]
\vspace{-0.09cm} 
\centering
{\adjincludegraphics[width=0.6\columnwidth,
trim={{0.06\width} {0.49\height} {0.09\width} {0.09\height}}
,clip]{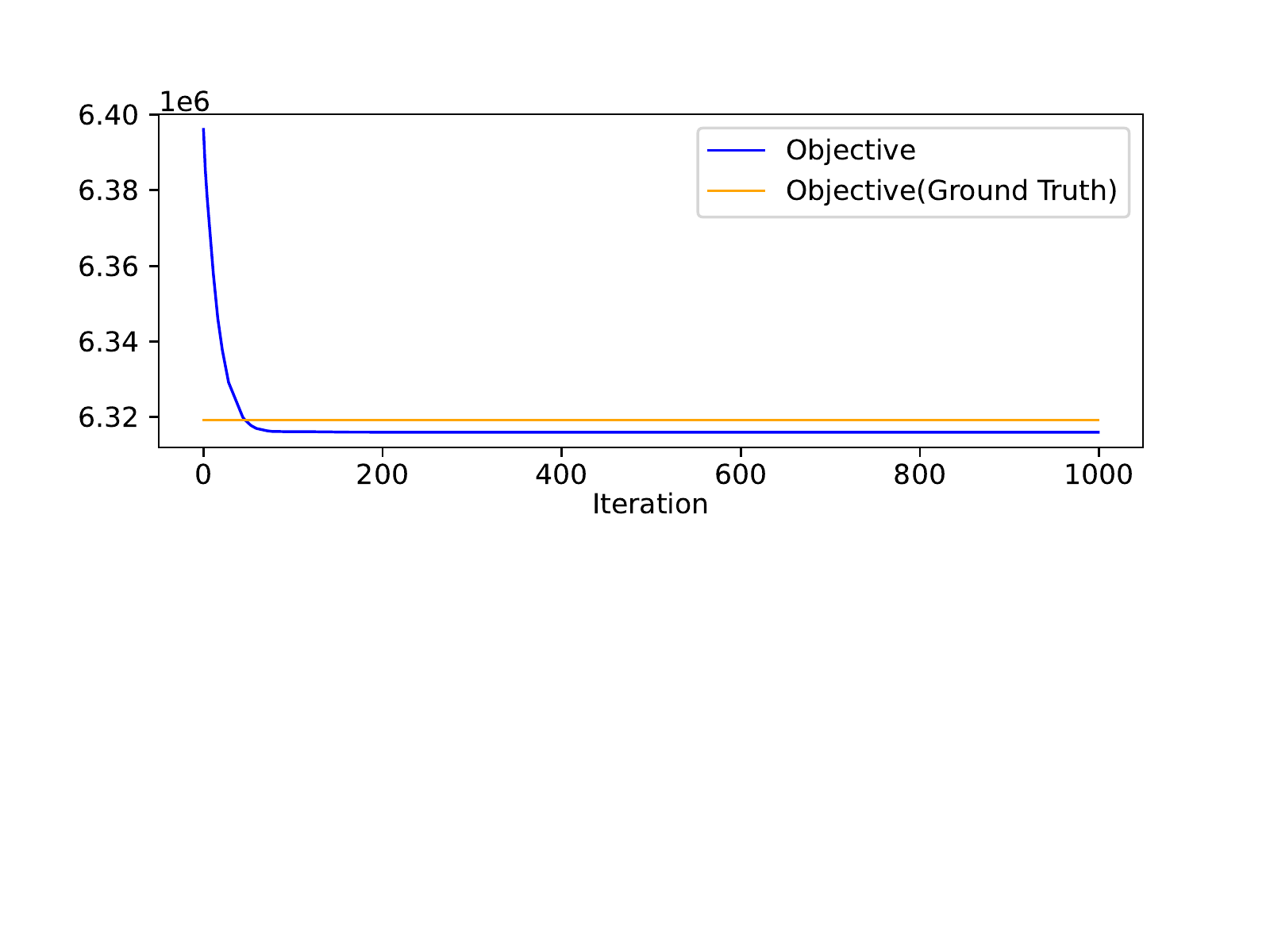}}
\caption[]{Convergence of areal density estimation for the simulated data.}
\label{fig:mcsim_convergence_Z}
\vspace{-0.09cm} 
\end{figure}

\about{convergence}
\fig{mcsim_convergence_Z} shows the convergence behaviour of the optimization of \eq{MaterialDecompositionEstimation}.
The final objective is slightly better than the ground truth objective after only about 50 iterations and then converges to its final value. This indicates likely convergence to near the global minimum, however, with possible slight overfitting. Note that the convergence takes significantly fewer iterations as compared to the nuisance parameter estimation.

\subsection{2D Reconstruction from Experimental Data}
\label{sec:2DExperimental}

\about{motivation}
This experiment is similar to the 2D Monte-Carlo simulated experiment of \sctn{2DSimulated}, but it uses measured data. We use a known, well-defined sample in order to demonstrate the quantitative accuracy of our method with experimentally measured data.

\about{setup}
The sample is a stack of uniform tantalum ($\n{Ta}$) and tungsten ($\n{W}$) plates. The plates have thicknesses of $\ell_{\n{Ta}} = 2.42 \, \n{mm}$ and $\ell_{\n{W}} = 1.75 \, \n{mm}$. We refer to this sample as the $\n{Ta}\n{W}$ sample. The $N_{\n{A}} = 2260$ TOF bins span an interval from $70.11 \, \n{\upmu s}$ up to $739.1 \, \n{\upmu s}$ which corresponds to an energy range of $115.0 \, \n{eV}$ down to $1.04 \, \n{eV}$, respectively. We use a cropped region of the detector of $N_{\n{p}} = 128 \!\times\! 128$ pixels.

\begin{figure}[htb]
\vspace{-0.09cm} 
\centering
{\adjincludegraphics[height=0.35\columnwidth,
trim={{0.10\width} {0.08\height} {0.10\width} {0.10\height}}
,clip]{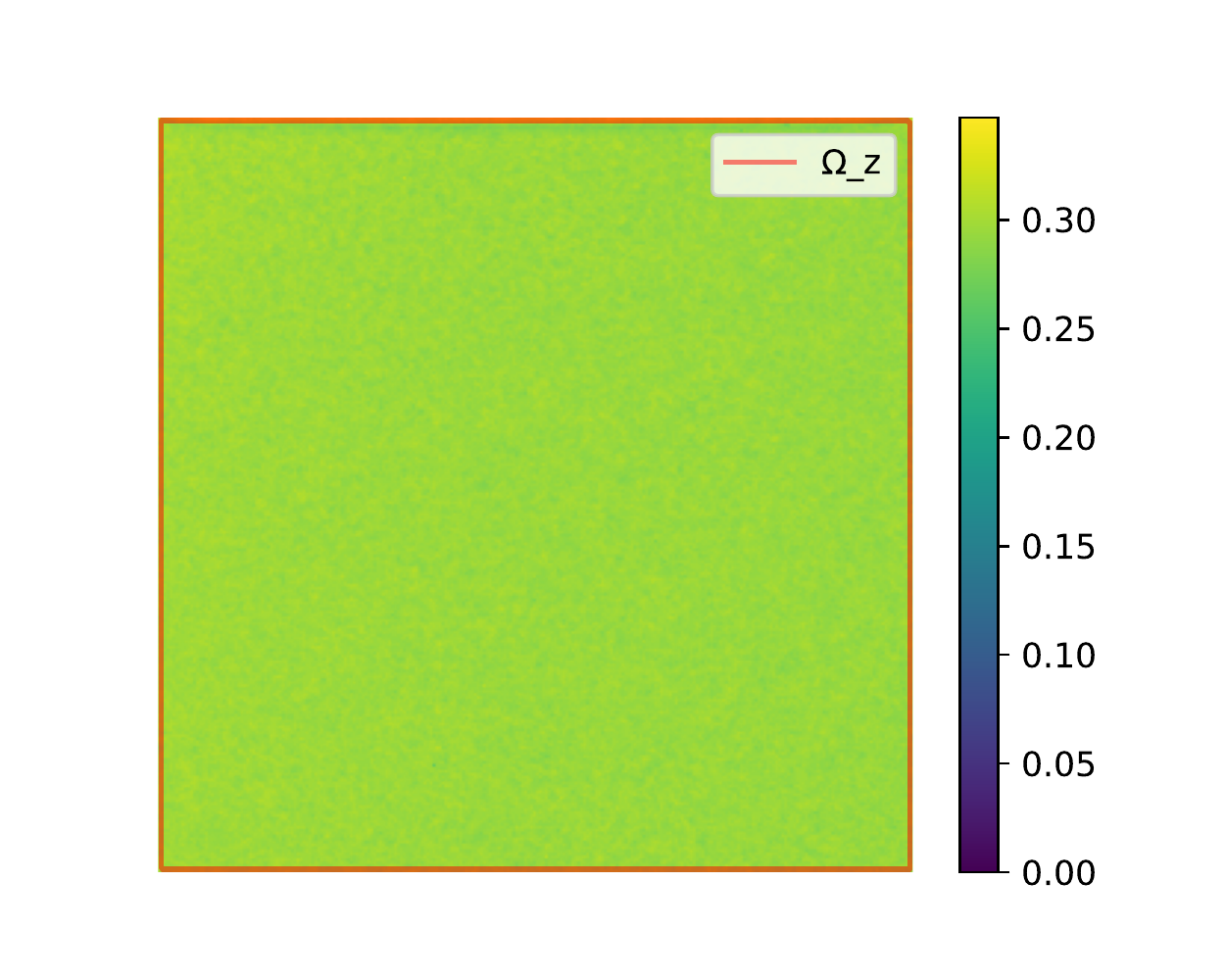}}
\caption[]{TOF integrated measurement ratio, $\sfrac{Y_{\n{s}} \mbb{1}}{Y_{\n{o}} \mbb{1}}$, of the $\n{Ta}\n{W}$ sample. Uniformly dense region, $\Omega_{\n{z}}$ (red line), open beam region, $\Omega_{\n{0}}$ does not exist for this measurement because the whole field of view is covered by the sample.}
\label{fig:coin_TaW_Cs1_div_Co1}
\vspace{-0.09cm} 
\end{figure}

\about{$\Omega_{\n{0}}$, $\Omega_{\n{z}}$}
\fig{coin_TaW_Cs1_div_Co1} shows the TOF integrated measurement ratio, $\sfrac{Y_{\n{s}} \mbb{1}}{Y_{\n{o}} \mbb{1}}$ of the sample. The sample covers the entire field of view and thus there is no set of pixels that is suitable for the open beam region $\Omega_{\n{0}}$. For this reason we choose the parameter $\beta = 0$
in \eq{NuisanceParamsEstimation}, effectively dropping the constraint of jointly solving \eq{y_s0_derivation} in the nuisance parameter estimation. Since the sample is assumed to be constant across the field of view, the set $\Omega_{\n{z}}$ is chosen to be the entire field of view. 

\begin{figure}[htb]
\vspace{-0.09cm} 
\centering
{\adjincludegraphics[width=1.0\columnwidth,
trim={{0.05\width} {0.04\height} {0.06\width} {0.12\height}}
,clip]{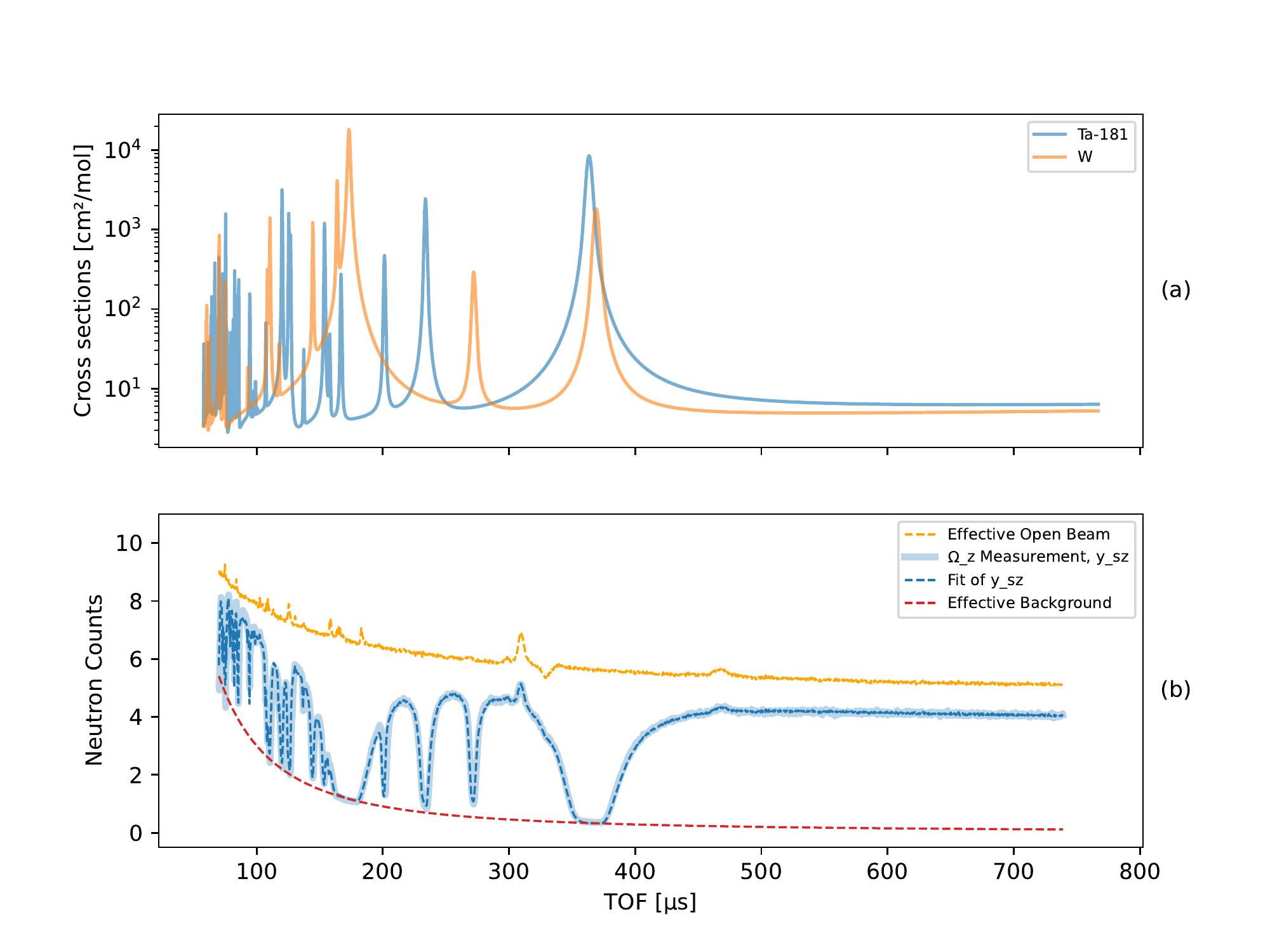}}
\caption[]{Nuisance parameter estimation fits of the $\n{Ta}\n{W}$ sample.\\
(a) Cross section dictionary, $D$, with $^{181}\n{Ta}$ isotope and Tungsten $\n{W}$ in natural abundances in units of $\n{cm^2/mol}$.\\
(b) Effective open beam (yellow dashed), $\Omega_{\n{z}}$ measurement $\mb{y}_{\n{sz}}$ (blue), fit of $\mb{y}_{\n{sz}}$ (blue dashed), effective background (red dashed).}
\label{fig:coin_TaW_fit_solo}
\vspace{-0.09cm} 
\end{figure}

\about{dictionary}
\fig{coin_TaW_fit_solo}(a) shows the cross section dictionary, $D$, used for this experiment. Since tantalum in natural abundance consists of practically $100 \, \%$ of the $^{181}\n{Ta}$ isotope, we use this isotope for the cross section dictionary. In contrast, natural tungsten consists of several isotopes, so for its cross section we use a proportionally weighted sum of the cross section of its isotopes, $^{182}\n{W}$, $^{183}\n{W}$, $^{184}\n{W}$, and $^{186}\n{W}$.

\about{fits}
\fig{coin_TaW_fit_solo}(b) shows the plots corresponding to the nuisance parameter estimation of \eq{NuisanceParamsEstimation}. Note that (compared to \fig{mcsim_prep_fit}), the fit of $\mb{y}_{\n{s0}}$ is not shown since there is no $\Omega_{\n{0}}$ region. Nevertheless, the optimization seems to result in a good fit and plausible spectra of the effective flux and background. 

\begin{table}[htb]
\vspace{-0.09cm} 
\centering
\caption{\label{tab:TaW_estimates} Nuisance parameter values for the $\n{Ta}\n{W}$ sample.}
\scalebox{0.92}{
    \begin{tabular}{@{}c|c@{}}
    \toprule
    Quantity & Estimates\\ 
    \midrule
    $\left[ \begin{matrix}\n{isotopes} \\ \hat{\mb{z}}^\top \, ^{[\n{mmol/cm^2}]} \end{matrix} \right]$
    & $\left[\begin{matrix} ^{181}\n{Ta} & \n{W} \\ 20.59 & 21.68  \end{matrix} \right]$ \\
    $\alpha_1, \ \ \alpha_2$ &  $0.483, \ \ 0.685$\\
    $\mbs{\uptheta}^\top$ &  $\left[\begin{matrix}29.9 & -56.1 & 5.39\end{matrix}\right]$\\
    \bottomrule
    \end{tabular}
}
\vspace{-0.09cm} 
\end{table}

\about{numerical values}
\tab{TaW_estimates} shows the numerical values of the nuisance parameters. Note that the $\alpha_2$ parameter is relatively different from $1$, indicating that the effects of the sample on the background are important to model.

\begin{figure}[htb]
\vspace{-0.09cm} 
\centering
{\adjincludegraphics[height=0.35\columnwidth,
trim={{0.12\width} {0.09\height} {0.12\width} {0.05\height}}
,clip]{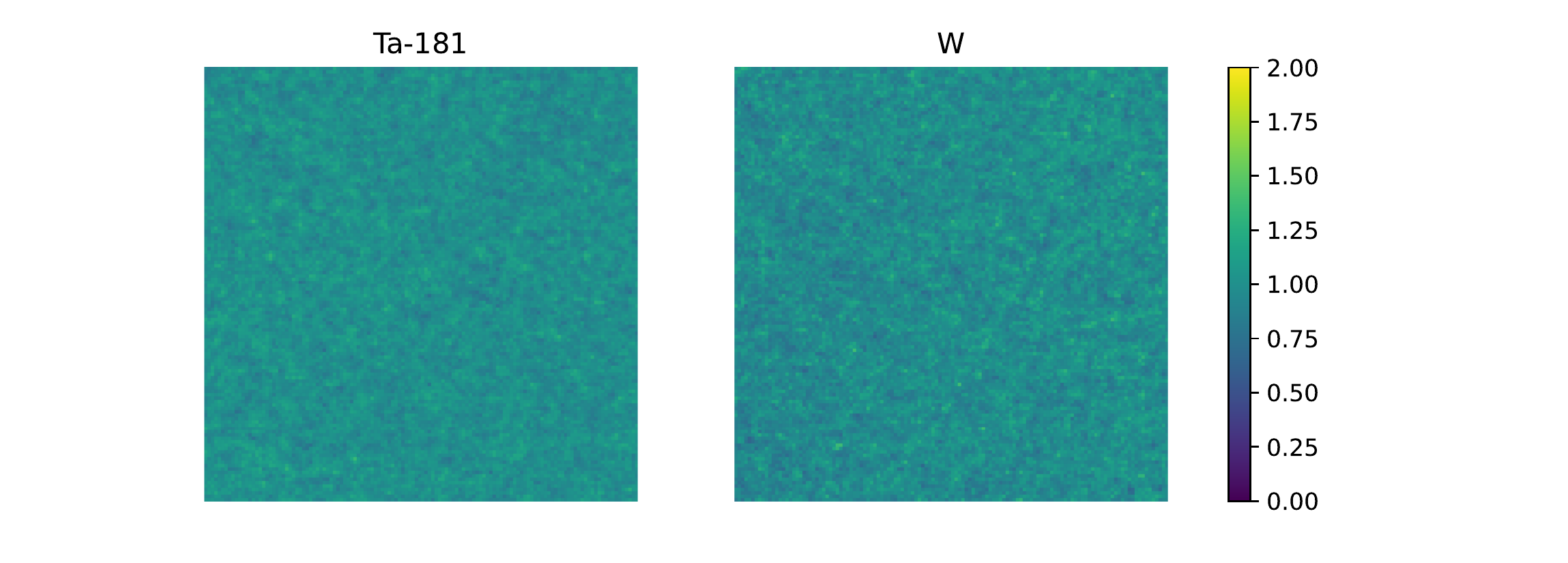}}
\caption{Normalized areal density estimates, $\hat{Z} \diag(\hat{\mb{z}})^{-1}$,  for the $\n{Ta}\n{W}$ sample [unitless].}
\label{fig:coin_TaW_Z_hat}
\vspace{-0.09cm} 
\end{figure}

\about{2D reconstruction}
\fig{coin_TaW_Z_hat} shows the areal density estimates, $\hat{Z} \diag(\mb{z})^{-1}$. 
We observe that, apart from some noise, the estimates are visually relatively constant as is expected from a spatially constant sample. 
The noise level appears to be similar to the simulated data of \sctn{2DSimulated}, however, it seems to be more spatially correlated, indicating possible detector cross talk.

\begin{table}[htb]
\vspace{-0.09cm} 
\centering
\caption{\label{tab:TaW_thickness} Mean and standard deviation of areal densities and thicknesses vs. ground truth thicknesses of the $\n{Ta}\n{W}$ sample.}
\scalebox{0.92}{
    \begin{tabular}{@{}l|ll@{}}
    \toprule
    \hspace{1cm} ${}_\text{Quantity} \setminus {}^\text{Isotopes}$    & $^{181}\n{Ta}$ & $\n{W}$  \\ 
    \midrule
    Areal Density Estimate $^{[\n{\frac{mmol}{cm^2}}]}$ & $20.56 \pm 1.59$ & $21.11 \pm 2.09$\\
    \hrulefill & \hrulefill & \hrulefill \\
    Ground Truth Thickness $^{[\n{cm}]}$         & $0.242$                     & $0.175$\\
    \textbf{Estimated Thickness} $^{[\n{cm}]}$   & $\mb{0.223} \pm \mb{0.017}$ & $\mb{0.197} \pm \mb{0.020}$\\
    \bottomrule
    \end{tabular}
}
\vspace{-0.09cm} 
\end{table}

\about{thickness estimates}
Since the $\n{Ta}\n{W}$ sample matches the idealized example from \sctn{ImagingSystem} we can use \eq{arealdensity_plate} and the areal density estimates, to make a thickness estimate for each of the plates. \tab{TaW_thickness} shows the mean and standard deviation computed from the areal density estimates, $\hat{Z}$, the resulting estimated thicknesses and the ground truth thicknesses of the plates. The estimated thicknesses are relatively close to the actual thicknesses with an error of less than $13\%$.

\begin{figure}[htb]
\vspace{-0.09cm} 
\centering
{\adjincludegraphics[width=1.0\columnwidth,
trim={{0.05\width} {0.0\height} {0.10\width} {0.52\height}}
,clip]{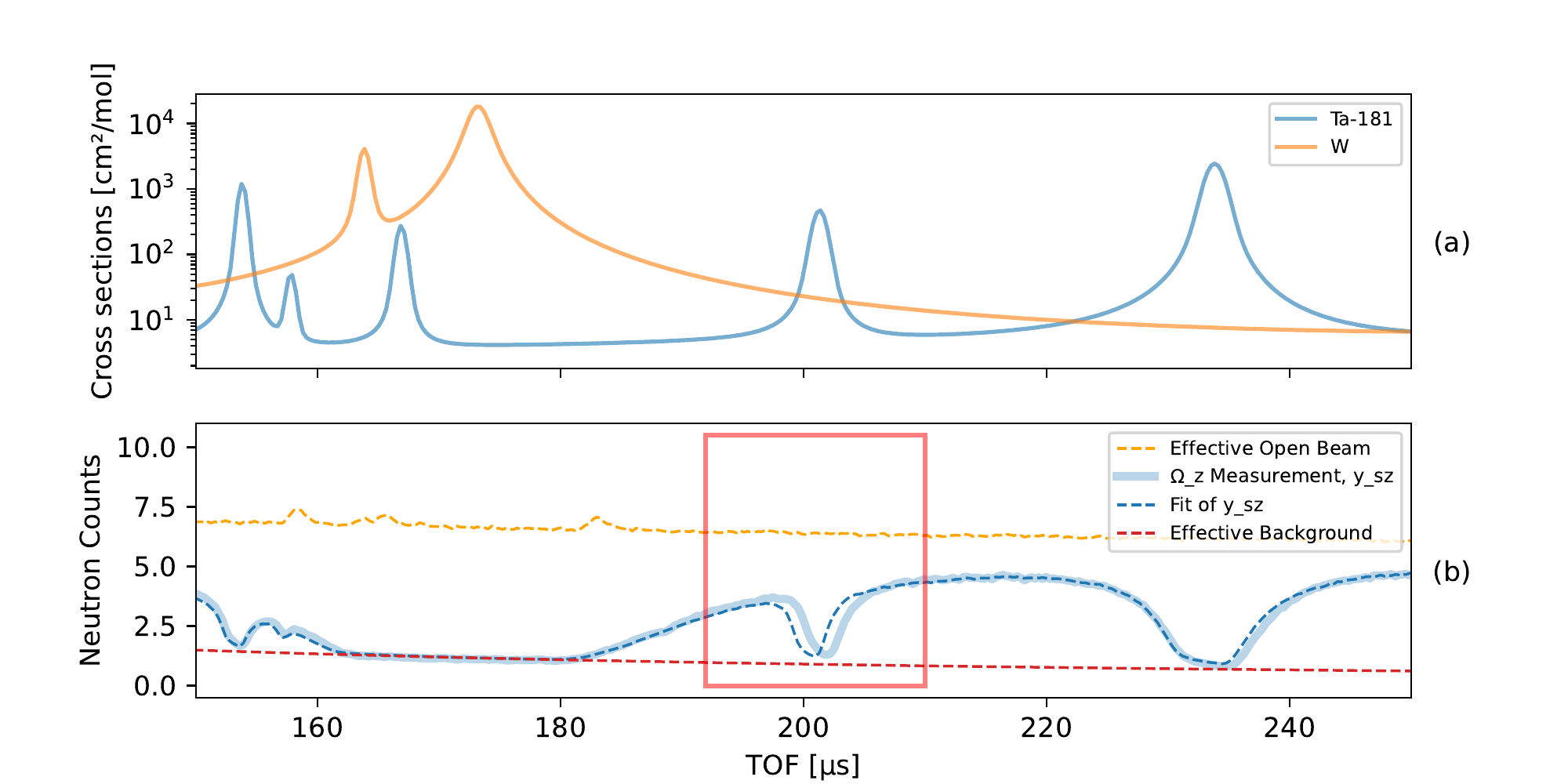}}
\caption{Nuisance parameter estimation fits for the $\n{Ta}\n{W}$ sample. The plot shows a cropped TOF region where the red box indicates likely error in the tabulated energy of a Ta resonance.
}
\label{fig:coin_TaW_fit_crop}
\vspace{-0.09cm} 
\end{figure}

\about{why not more accurate?}
We believe that the slight inaccuracies in \tab{TaW_estimates} may be primarily inaccuracies in the assumed values of $R$ and $D$. We rely on previous measurements by~\cite{lynn2002neutron} for the parameters of $R$, which may not be sufficiently accurate. In addition, for the cross sections, $D$, we use tabulated data from ENDF/B-VIII.0\cite{conlin2018release}, which is empirically compiled and may not be sufficiently accurate for our experiment. In order to support our conjecture, \fig{coin_TaW_fit_crop} highlights a region (red box) in the spectrum where there are likely errors in $D$. In this region, the fitted resonance energy (blue dashed) in $^{181}\n{Ta}$ near $200 \, \n{\upmu s}$ has a slightly lower TOF than the measured resonance (blue solid). This type of mismatch can significantly affect the density estimates. There are likely other errors in the cross sections, which may however not be as obvious to spot as this one.

\subsection{3D Reconstruction from Experimental Data}
\label{sec:3DExperimental}

\about{goal}
In this experiment we demonstrate the tomographic reconstruction extension that is described in \sctn{tomography} on measured experimental data of a transmutation nuclear fuel pellet sample. We will also compute average volumetric mass densities and validate them through independent mass spectroscopy measurements.

\begin{figure}[ht]
\vspace{-0.09cm} 
    \centering
    \begin{minipage}{0.47\columnwidth}
        \centering
        {\adjincludegraphics[height=1.3\columnwidth,
        trim={{0.0\width} {0.09\height} {0.0\width} {0.11\height}}
        ,clip]{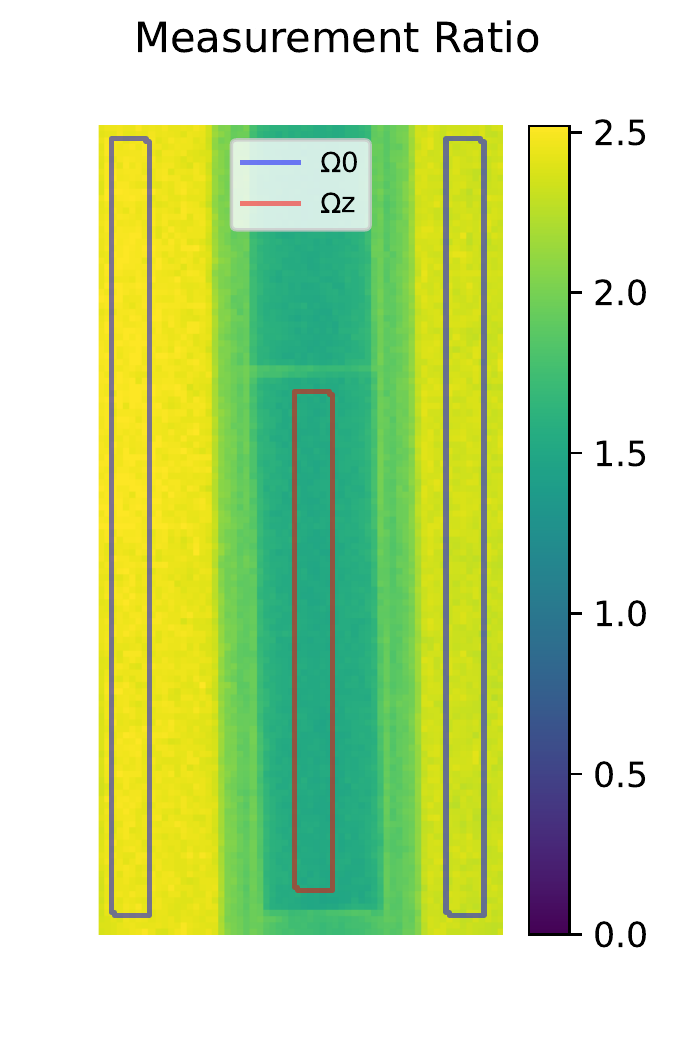}}
        \caption{TOF integrated measurement ratio, $\sfrac{Y_{\n{s}}^{(k)} \mbb{1}}{Y_{\n{o}} \mbb{1}}$, of the fuel pellet sample of a single view. Open beam region, $\Omega_{\n{0}}$ (blue), and uniformly dense region, $\Omega_{\n{z}}$ (red).}
        \label{fig:upuzr_regions}
    \end{minipage}\hfill
    \begin{minipage}{0.47\columnwidth}
        \centering
        {\adjincludegraphics[height=1.3\columnwidth,
        trim={{0.0\width} {0.11\height} {0.0\width} {0.08\height}}
        ,clip]{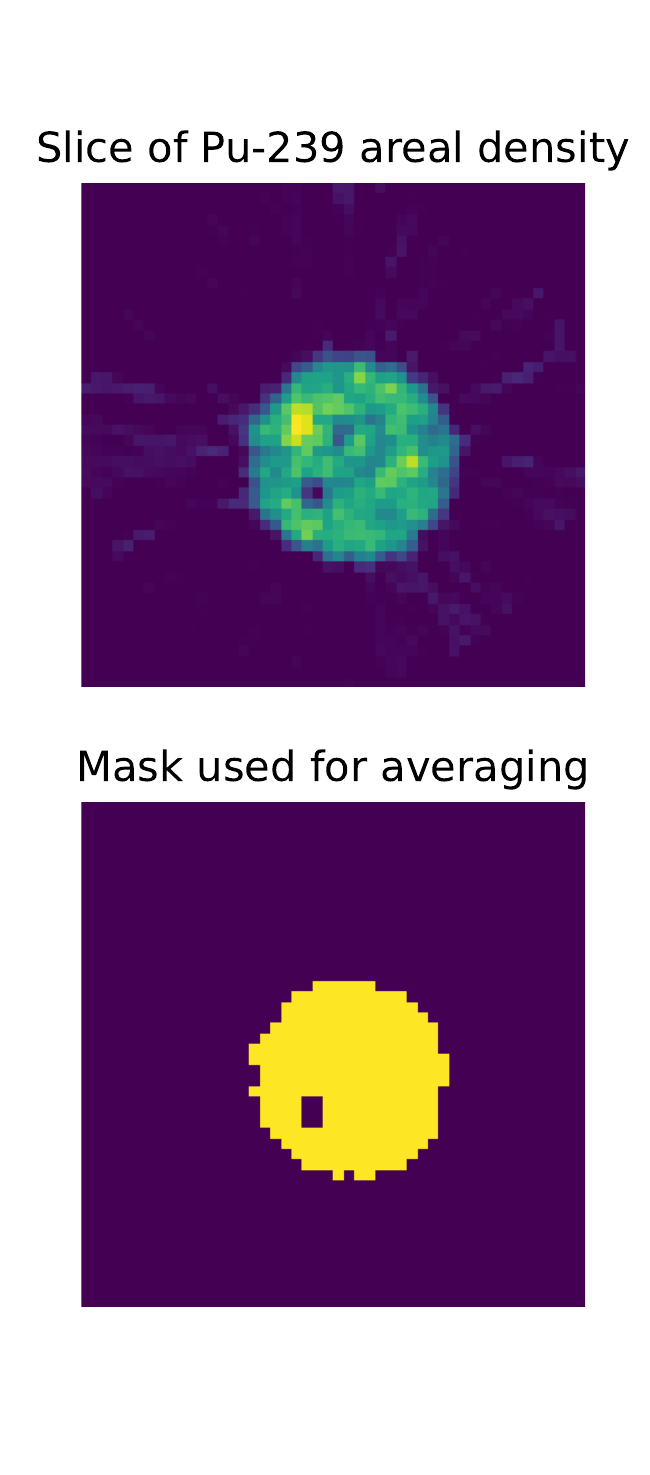}}
        \caption[]{Slice of the volumetric density, $\hat{X}$, of $^{239}\n{Pu}$ and corresponding mask generated through thresholding and used to determine average volumetric densities, $\hat{\mb{x}}$.}
        \label{fig:upuzr_x_mask}
    \end{minipage}
\vspace{-0.09cm} 
\end{figure}

\begin{figure}[htb]
\vspace{-0.09cm} 
\centering
{\adjincludegraphics[width=1.0\columnwidth,
trim={{0.05\width} {0.05\height} {0.05\width} {0.10\height}}
,clip]{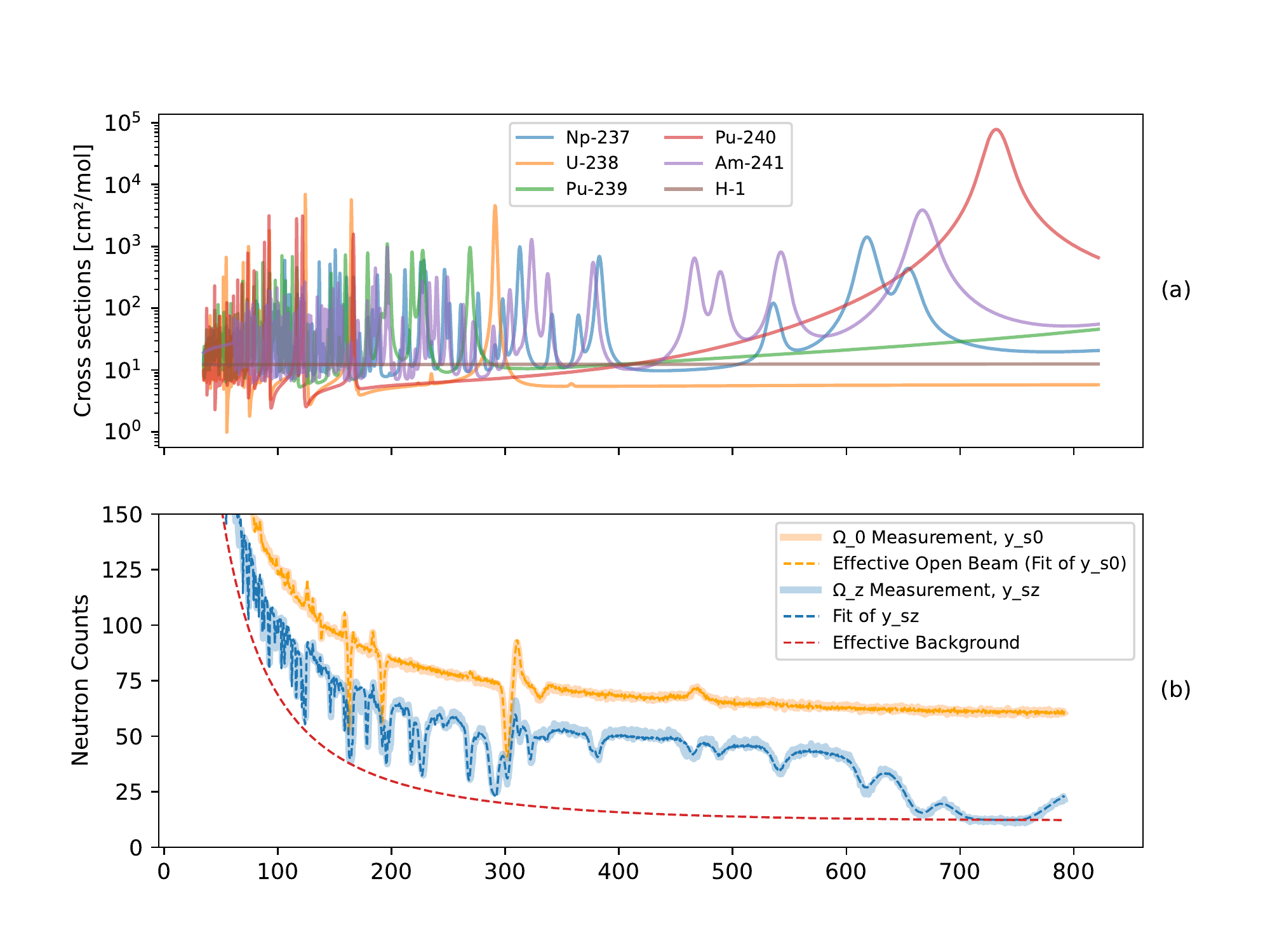}}
\caption[]{Nuisance parameter estimation fits of the fuel pellet sample.\\
(a) Cross section dictionary, $D$, with $^{237}\n{Np}$, $^{238}\n{U}$, $^{239}\n{Pu}$, $^{240}\n{Pu}$, $^{241}\n{Am}$, and $^{1}\n{H}$ isotopes in units of $\n{cm^2/mol}$.\\
(b) $\Omega_{\n{0}}$ measurement $\mb{y}_{\n{s0}}$ (yellow), effective open beam (yellow dashed), $\Omega_{\n{z}}$ measurement $\mb{y}_{\n{sz}}$ (blue), fit of $\mb{y}_{\n{sz}}$ (blue dashed), effective background (red dashed).}
\label{fig:upuzr_fit}
\vspace{-0.09cm} 
\end{figure}

\begin{figure*}[ht]
\vspace{-0.09cm} 
\centering
{\adjincludegraphics[width=1.6\columnwidth,
trim={{0.12\width} {0.08\height} {0.18\width} {0.02\height}}
,clip]{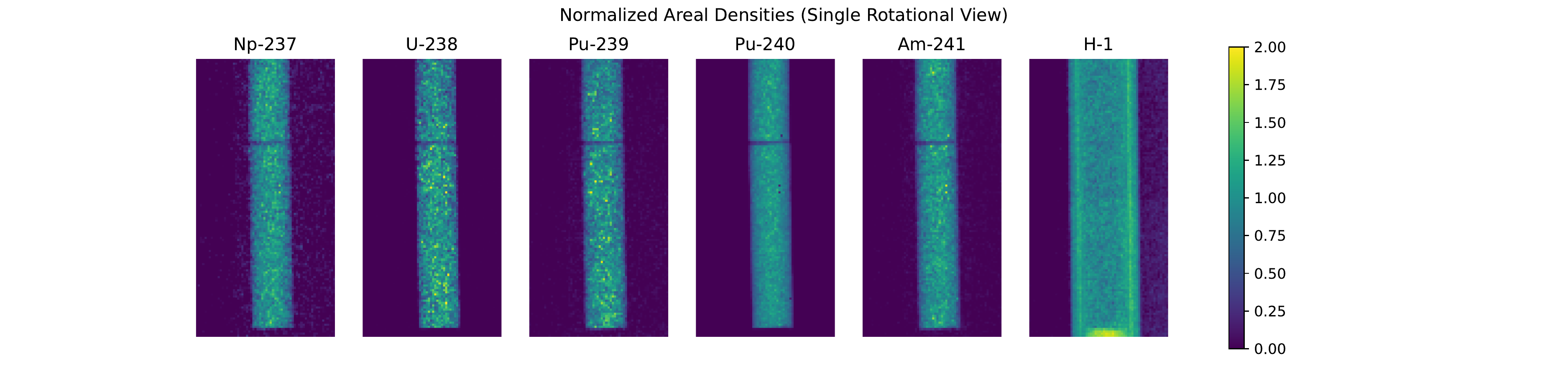}}
\caption[]{Single view of the normalized areal density results for the fuel pellet sample, $\hat{Z}^{(k)} \diag(\hat{\mb{z}})^{-1}$ [unitless].}
\label{fig:upuzr_Z}
\vspace{-0.09cm} 
\end{figure*}

\begin{figure*}[ht]
\vspace{-0.09cm} 
\centering
{\adjincludegraphics[width=1.6\columnwidth,
trim={{0.12\width} {0.08\height} {0.18\width} {0.02\height}}
,clip]{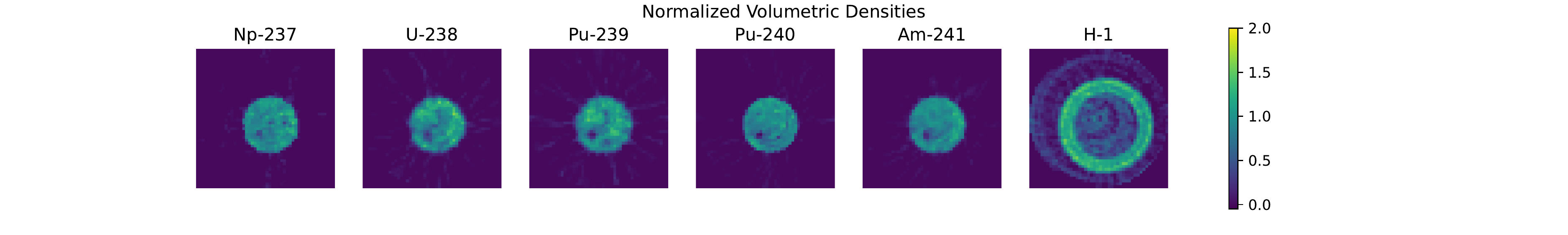}}
\caption[]{Slice of the normalized volumetric density results for the fuel pellet sample, $\hat{X} \diag(\hat{\mb{x}})^{-1}$, [unitless]. ($^{1}\n{H}$ is normalized as $\hat{X}_{*,i} / (3 \hat{\mb{x}}_i)$) }
\label{fig:upuzr_X}
\vspace{-0.09cm} 
\end{figure*}

\about{sample}
\fig{upuzr_regions} shows the TOF integrated measurement ratio, $\sfrac{Y_{\n{s}}^{(k)} \mbb{1}}{Y_{\n{o}} \mbb{1}}$, of a single rotational view. The cylindrical nuclear fuel pellet is surrounded by a double wall steel cladding. We choose the regions surrounding the sample as the $\Omega_{\n{0}}$ region and a thin axial region the center of the sample as the $\Omega_{\n{z}}$ region in the radiograph selected for the nuisance parameter estimation.

\about{isotopes}
\fig{upuzr_fit} shows the cross section dictionary, $D$. We assume that the sample contains the resonant isotopes $^{237}\n{Np}$, $^{238}\n{U}$, $^{239}\n{Pu}$, $^{240}\n{Pu}$, $^{241}\n{Am}$. We also assume there are non-resonant materials such as for example $\n{Zr}$ in the fuel and steel in the double wall cladding. For this reason, the cross section dictionary, $D$, not only includes the resonant isotopic cross sections but also the mostly constant $^{1}\n{H}$ (hydrogen) cross section. The reconstructed areal density associated with $^{1}\n{H}$ is interpreted as hydrogen-equivalent areal density, regardless whether the sample contains any hydrogen.

\about{other stats}
The sample is imaged at $N_{\n{r}} = 101$ different rotational views. The $N_{\n{A}} = 2440$ TOF bins span an interval from $47.6 \, \n{\upmu s}$ up to $791.7 \, \n{\upmu s}$ which corresponds to an energy range of $249.8 \, \n{eV}$ down to $0.902 \, \n{eV}$, respectively. We use a cropped region of the detector of $N_{\n{p}} = 256 \!\times\! 512$ pixels. We applied a $4 \!\times$ binning because the measurements suffer from severe defects such as many malfunctioning detector rows and columns. We understand that in general binning may not be the best way to remove such defects, however, for simplicity sake we deemed this procedure appropriate.

\about{nuisance parameter fits}
\fig{upuzr_fit}(b) shows the fits resulting from the nuisance parameter estimation of one single view. The good fits of the respective measurements indicate successful estimation of the parameters. \tab{fuel_pellet} lists the estimated parameters.

\about{areal density recon}
\fig{upuzr_Z} shows the normalized areal density reconstruction, $\hat{Z}^{(k)} \diag(\hat{\mb{z}})^{-1}$, of a single view from the material decomposition. Each of the resonant isotopes, $^{237}\n{Np}$, $^{238}\n{U}$, $^{239}\n{Pu}$, $^{240}\n{Pu}$, $^{241}\n{Am}$, clearly resolves the cylindrical fuel pellet while the $^{1}\n{H}$ equivalent density is mostly concentrated in the cladding.

\begin{table}[t]
\vspace{-0.09cm} 
\centering
\caption{\label{tab:fuel_pellet} Nuisance parameter values for the fuel pellet sample.}
\scalebox{0.92}{
    \begin{tabular}{@{}c|c@{}}
    \toprule
    Quantity & Estimates\\ 
    \midrule
    $\left[ \begin{matrix} \n{isotopes} \\ \hat{\mb{z}}^\top \, ^{[\n{mmol/cm^2}]} \end{matrix} \right]$
    & $\left[\begin{matrix} ^{237}\n{Np} & ^{238}\n{U} & ^{239}\n{Pu} & ^{240}\n{Pu} & ^{241}\n{Am} & ^{1}\n{H} \\ 0.57 & 12.9 & 2.89 & 0.738 & 0.584 & 23.8 \end{matrix} \right]$ \\
    $\hat{\alpha}_1, \ \ \hat{\alpha}_2$ &  $2.39, \ \ 1.02$\\
    $\hat{\mbs{\uptheta}}^\top$ &  $\begin{matrix}[112.3 & -39.0 & 15.1]\end{matrix}$\\
    \bottomrule
    \end{tabular}
}
\vspace{-0.09cm} 
\end{table}

\begin{table}[t]
\vspace{-0.09cm} 
\centering
\caption{\label{tab:fuel_pellet_2} Average values of volumetric densities, $\hat{\mb{x}}$, of the fuel pellet sample in units of ${[\n{mmol/cm^3}]}$.}
\scalebox{0.92}{
    \begin{tabular}{@{}c|c@{}}
    \toprule
    Quantity & Estimates\\ 
    \midrule
    $\left[ \begin{matrix} \n{isotopes} \\ \hat{\mb{x}}^\top  \end{matrix} \right]$
    & $\left[\begin{matrix} ^{237}\n{Np} & ^{238}\n{U} & ^{239}\n{Pu} & ^{240}\n{Pu} & ^{241}\n{Am} & ^{1}\n{H} \\ 1.394 & 37.87 & 8.369 & 1.814 & 1.409 & 55.7\\  \end{matrix} \right]$\\
    \bottomrule
    \end{tabular}
}
\vspace{-0.09cm} 
\end{table}

\begin{table}[t]
\vspace{-0.09cm} 
\centering
\caption{\label{tab:fuel_pellet_3} Mass density estimates the fuel pellet sample compared to mass spectrometry~\cite{vogel2018neutron} (ground truth) in units of $\n{g/cm^3}$.}
\scalebox{0.92}{
    \begin{tabular}{@{}l|lllll@{}}
    \toprule
    ${}_\text{Mass Density} \setminus {}^\text{Isotopes}$
    & $^{237}\n{Np}$ & $^{238}\n{U}$ & $^{239}\n{Pu}$ & $^{240}\n{Pu}$ & $^{241}\n{Am}$\\ 
    \midrule
    Ground Truth & 0.34 & 9.01 & 2.34 & 0.37 & 0.32\\
    \textbf{Estimated} & \textbf{0.331} & \textbf{9.014} & \textbf{2.000} & \textbf{0.435} & \textbf{0.340}\\
    \bottomrule
    \end{tabular}
}
\vspace{-0.09cm} 
\end{table}

\about{volumetric densities}
\fig{upuzr_X} shows a single axial slice of the normalized volumetric density reconstruction, $\hat{X} \diag(\hat{\mb{x}})^{-1}$, from \eq{ct}. We computed the vector $\hat{\mb{x}} \in \mbb{R}^{N_{\n{m}}}$ as the average estimated volumetric density in the mask shown in \fig{upuzr_x_mask}. The the values of $\hat{\mb{x}}$ are shown in \tab{fuel_pellet_2}. In the displayed slice the clear separation between fuel and cladding is visible. Also, there is a void visible in the 7-o'clock position in all resonant isotope maps, indicating what is probably a void in the sample.

\about{mass densities}
Finally, \tab{fuel_pellet_3} shows the average reconstructed mass density estimates, $\hat{\mbs{\uprho}} \, ^{[\n{g/cm^3}]}$, computed from the volumetric densities, $\hat{\mb{x}} \, ^{[\n{mol}/\n{cm}^3]}$ and the relationship in \eq{volumetricdensity_plate}.
The table also lists the values of ground truth measurements obtained using mass spectroscopy.
Note that the reconstructed estimates are very close to the ground truth values which provides strong experimental validation of our method.

\section{Conclusions}
\label{sec:Conclusion}

We present a novel method for both 2D and 3D reconstruction of material areal and volumetric density from TOF neutron images.
Our method, TRINIDI, accounts for a number of important measurement effects including spatially varying background, non-uniform flux, finite pulse duration, and Poisson counting noise. Especially, when including the finite pulse duration in the forward model, the problem cannot be linearized in the same fashion that is common with Beer-Law attenuation models.
TRINIDI is based on a 2-step process in which we first estimate nuisance parameters associated with background and flux, and we then reconstruct the unknown isotopic areal density. 
For the 3D case, we then perform tomographic reconstruction of the views to form a 3D volumetric density estimate.
Both our simulated and experimental results indicate that TRINIDI can reconstruct quantitatively accurate estimates in both 2D and 3D.
Notably, comparisons of our 3D material decomposed reconstructions for nuclear fuel pellets are accurate when compared to mass spectroscopy measurements.

\begin{appendices}
\section{Resolution Function}
\label{app:ResolutionFunction}

\begin{figure}[htb]
\vspace{-0.09cm} 
\centering
{\adjincludegraphics[width=1.0\columnwidth,
trim={{0.09\width} {0.02\height} {0.07\width} {0.07\height}}
,clip]{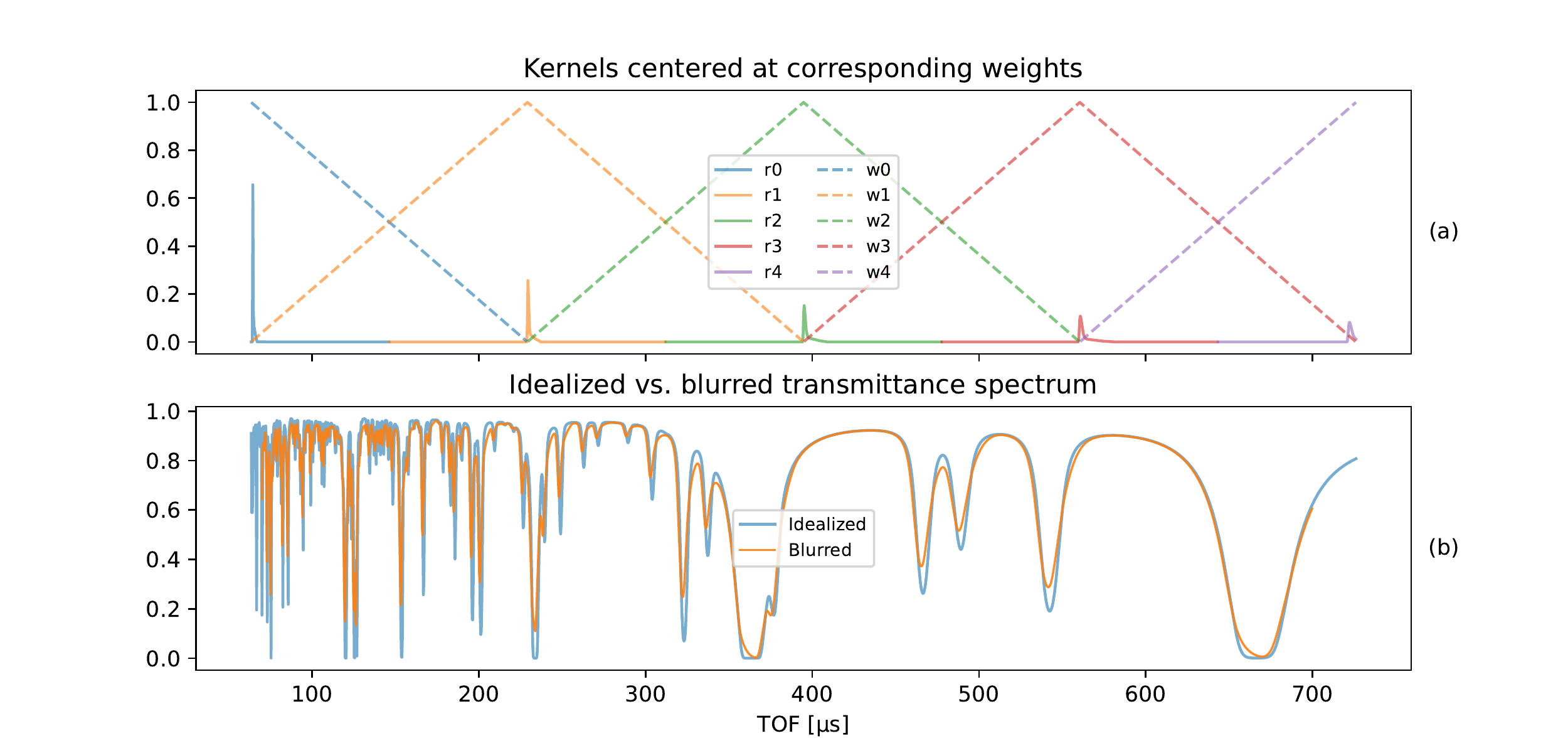}}
\caption[]{Illustration of resolution function with $K=5$ components applied to transmission spectrum. \\
(a) Dashed lines: $\mb{w}^k$ weights. Solid lines: $\mb{r}^k$ blurring kernels centered at TOF where they are most active.\\
(b) Blue line: non-blurred transmission spectrum. Orange line: blurred transmission spectrum using the kernels above.
}
\label{fig:resolution_demo}
\vspace{-0.09cm} 
\end{figure}

The resolution operator, $R$, models the uncertainty of emission time of the neutrons given their known neutron energy. Equivalently, it models the conditional probability of TOA's given the neutron's TOF's, stated more precisely in \eq{resolution_def}.
Since the pulse shape changes with every neutron energy, we choose to approximate the resolution operator using a weighted sum of $K$ convolution operations, so that
\begin{equation}
    \label{eq:R_def}
    R = \sum_{k=0}^{K-1} R^k W^k \ .
\end{equation}
The $R^k$ matrix corresponds to the convolution operation with kernel $\mb{r}^k$, which in turn corresponds to the pulse shape at the TOF bin with index $\lfloor k\frac{N_{\n{F}}-1}{K-1} \rfloor$.
The $W^k$ matrices are diagonal weighting matrices with diagonals $\mb{w}^k$ which blend together the blurring operators, $R^k$.

\fig{resolution_demo} illustrates the application of the resolution operator with $K = 5$ components in a typical scenario. In \fig{resolution_demo}(a) the dashed lines are the diagonals, $\mb{w}^k$ and the solid lines are the corresponding kernels, $\mb{r}^k$ centered at the TOF where they are most active. The blue line in \fig{resolution_demo}(b) shows a typical non-blurred transmission spectrum, $\exp(-\mb{z}^\top D)$. The orange line shows the corresponding blurred transmission spectrum, $\exp(-\mb{z}^\top D)R$. Note that the blurring is weaker in the low-TOF region and stronger in the high-TOF region and that black resonances in the non-blurred signal are mostly not opaque after the blurring.

Next we show that $\mbb{1}^\top R = \mbb{1}^\top$. First, we select the weights so every TOA bin they sum to $1$, i.e.
\begin{equation}
    \label{eq:sum_Wk}
    \sum_{k=0}^{K-1} W^k = I \, .
\end{equation}
Second, we assume that $\sum_j \mb{r}^k_{j} = 1$ so that every emitted neutron is measured as some TOA bin. As explained in \sctn{ForwardModel} this is achieved by choosing the number of rows of $R^k$ to be sufficiently large so that 
\begin{equation*}
    (\forall k) \ \sum_{i} R^k_{i,j} = \sum_{i} \mb{r}^k_{j - i} =  \sum_{i} \mb{r}^k_{j} = 1 \ ,  
\end{equation*}
so we have that
\begin{equation}
    \label{eq:sum_Rk}
    (\forall k) \ \mbb{1}^\top R^k = \mbb{1}^\top \, .
\end{equation}
Finally, using \eq{R_def}, \eq{sum_Wk}, and \eq{sum_Rk} we see that
\begin{align}
\label{eq:1R1}
\mbb{1}^\top R 
    &= \mbb{1}^\top \left( \sum_{k=0}^{K-1} R^k W^k \right)
    = \sum_{k=0}^{K-1}  \left( \mbb{1}^\top R^k \right) W^k  \nonumber\\ 
    &=         \sum_{k=0}^{K-1}  \mbb{1}^\top W^k
    = \mbb{1}^\top  \left( \sum_{k=0}^{K-1}   W^k \right) 
    = \mbb{1}^\top  I = \mbb{1}^\top \,.
\end{align}


\section{Spectral Basis Matrix of the Background}
\label{app:P-Matrix}

\about{$P$ basis functions}
In this work, we choose the background basis to be
\begin{equation}
P_{n,k} =
\frac{1}{a_n }
\left(
\log \left[ k \Delta + k_0 \right]
\right)^n \, ,
\label{eq:DefinitionOfP}
\end{equation}
where
$
a_n = 
\sqrt{
\sum_{k=0}^{N_{\n{A}}-1}
[
\left(
\log \left[ k \Delta + k_0 \right]
\right)^n 
]^2} 
$
scales the rows of $P$ so they have unit norm
and $0\leq n < N_{\n{b}}$, $0\leq k < N_{\n{A}}$, and we have found that scalar $\Delta = \frac{\n{e}^1 - \n{e}^{-1}}{N_{\n{A}} - 1}$ and the offset $k_0 = \n{e}^{-1}$ work well across a range of cases.

\end{appendices}



\bibliographystyle{IEEEbib}
\bibliography{refs}

\vfill

\end{document}